\documentclass[%
 aip,
 amsmath,amssymb,
 reprint,%
]{revtex4-1}

\usepackage{graphicx}
\usepackage{dcolumn}
\usepackage{bm}

\usepackage[utf8]{inputenc}
\usepackage[T1]{fontenc}
\usepackage{mathptmx}
\usepackage[dvipsnames]{xcolor}

\newcommand*\aap{A\&A}

\newcommand*\apjl{ApJ}

\newcommand*\apss{Ap\&SS}

\newcommand*\mnras{MNRAS}

\newcommand*\ssr{Space~Sci.~Rev.}

\begin{document}

\preprint{AIP/123-QED}

\title[Ion-neutral decoupling in the Kelvin-Helmholtz instability]{Ion-neutral decoupling in the nonlinear Kelvin--Helmholtz instability: Case of field-aligned flow}

\author{A. Hillier}
 \email{a.s.hillier@exeter.ac.uk}
\affiliation{ 
Department of Mathematics, CEMPS, University of Exeter, Exeter EX4 4QF U.K. 
}%

\date{\today}

\begin{abstract}
The nonlinear magnetic Kelvin-Helmholtz instability (KHi), and the turbulence it creates, appears in many astrophysical systems. This includes those systems where the local plasma conditions are such that the plasma is not fully ionised, for example in the lower solar atmosphere and molecular clouds. In a partially ionised system, the fluids couple via collisions which occur at characteristic frequencies, therefore neutral and plasma species become decoupled for sufficiently high-frequency dynamics. Here we present high-resolution 2D two-fluid simulations of the nonlinear KHi for a system that traverses the dynamic scales between decoupled fluids and coupled dynamics. We discover some interesting phenomena, including the presence of a density coupling that is independent of the velocity coupling. Using these simulations we analyse the heating rate, and two regimes appear. The first is a regime where the neutral flow is decoupled from the magnetic field that is characterised with a constant heating rate, then at larger scales the strong coupling approximation holds and the heating rate. At large scales with the KHi layer width to the $-2$ power. There is an energy cascade in the simulation, but the nature of the frictional heating means the heating rate is determined by the largest scale of the turbulent motions, a fact that has consequences for understanding turbulent dissipation in multi-fluid systems.
\end{abstract}

\maketitle

\section{Introduction}

The magnetohydrodynamic (MHD) Kelvin-Helmholtz instability (KHi) is a fundamental instability of magnetized plasma. It occurs in systems that have unstable velocity shears, resulting in the formation of vortices.
This instability has been observed in many astrophysical systems including the flanks of coronal mass ejections \citep[e.g.][]{FOULLON2011}, the flanks of the Earth's magnetosphere \citep[e.g.][]{HASE2004} and in solar prominences \citep[e.g.][]{BERG2010, BERG2017, HILL2018}.

The classic hydrodynamic (HD) KHi is the instability of a discontinuous velocity shear, and is unstable for any difference in velocity across the flow discontinuity.
In systems where magnetic fields are present, they can play a key role in the suppression of the instability \citep{CHAN1961} as the bending of magnetic fields to create vortices results in a magnetic tension force that works to suppress the vortex growth.
A necessary condition for this instability to form in generic field-aligned flows (including discontinuous flows) is that the difference between the maximum and minimum velocity must exceed twice the minimum Alfv\'{e}n speed \citep{HUGHES2001}.

Due to the low temperatures of many astrophysical systems (e.g. stellar atmospheres and protoplanetary disks), there is insufficient energy to fully ionise the plasma. As a result the majority of the fluid is composed of neutral species.
Meaning that even though magnetic fields can play an important dynamic role, the plasma itself is composed mostly of species that do not directly feel the magnetic field.
In this regime, there is a drift in the velocity between ions and neutrals because of the different forces the different species feel \citep[e.g.][]{KHOM2014, KHOM2017}.
These physical processes are found to be important in a wide range of astrophysical plasmas \citep{BALLE2018}.

This drift between the species means that when they collide, momentum is transferred and over time the drift between the species will reduce resulting in the two fluids flowing together.
Because this drift velocity between the species can result in frictional heating, the coupling process can be dynamically important for energy dissipation \citep[e.g][]{KHOM2012}.
Single fluid approximations, using ambipolar diffusion in a modified Ohm's law, can be applied when the dynamic frequency is much smaller than the collision frequencies \citep[e.g.][]{BRAG1965}. However, when this is not the case, multifluid modelling is necessary \citep[e.g.][]{HILL2016}.

Investigations into the partially ionised plasma effects on the KHi have been performed for both linear and nonlinear settings.
Investigations have been performed into the linear stability condition for the KHi in the interstellar medium, assuming an incompressible flow with constant density \citep{WATSON2004}. The key result found in this study is that for much of the parameter range of interest, the neutrals would be unstable but the ions would be held in place by the magnetic field \citep{WATSON2004}.
The linear stability conditions for the two-fluid KHi in solar prominences show that in the incompressible limit instability is present for any velocity shear \citep{SOLER2012}. However, the inclusion of compressibility meant that the onset velocity shear value became heavily dependent on the parameters used, especially the density contrast and collision frequency \citep{SOLER2012}. 
Applying these results to the case of the KHi driven by prominence plumes \citep[e.g.][]{BERG2010}, it was found that sub-Alfvénic flow velocities could trigger the instability as a result of the finite ion-neutral coupling \citep{SOLER2012}.
Further studies have been performed for the linear and nonlinear development of the KHi in a partially-ionised dusty plasma \citep{BIRK2002}, and into the effect of a finite boundary \citep{SHADMEHRI2008}. 

Nonlinear simulations, performed for the case of a very low ionisation degree which is appropriate for molecular clouds have been performed for both single fluid modelling \citep{JONES2011} and for multifluid (neutral, ion, electron and dust) modelling \citep{JONES2012} . 
Significant differences from reference MHD cases were found, with the ion-neutral drift resulting in a significantly weaker wind-up of the magnetic field.

In this paper, we present a study of the nonlinear regime to investigate the physics of the KHi in partially ionised plasma. We focus is on what happens for the nonlinear KHi at different dynamic frequencies ($\nu_D$) when compared with the collisions of the neutrals on the ions ($\nu_{\rm np}$) and the ions on the neutrals ($\nu_{\rm pn}$) and how this influences the turbulent structures and dissipation at different scales.

\section{Methods}

\subsection{Partially Ionised Plasma MHD}

In this study we investigate the dynamics of a neutral fluid and a charge-neutral fully-collisionally-coupled ion electron plasma. 
The nondimensional equations solved for the evolution of the neutral hydrogen fluid are:
\begin{align}
\frac{\partial\rho_{\rm n}}{\partial t}+\nabla\cdot(\rho_{\rm n}\mathbf{v}_{\rm n})=&0, \\
\frac{\partial}{\partial t}(\rho_{\rm n}\mathbf{v}_{\rm n})+\nabla\cdot(\rho_{\rm n}\mathbf{v}_{\rm n}\mathbf{v}_{\rm n}+P_{\rm n}\mathbf{I})=&-\alpha_{\rm c}\rho_{\rm n}\rho_{\rm p}(\mathbf{v}_{\rm n}-\mathbf{v}_{\rm p}),\\
\frac{\partial e_{\rm n}}{\partial t}+\nabla\cdot[\mathbf{v}_{\rm n}(e_{\rm n}+P_{\rm n})]=-\alpha_{\rm c}\rho_{\rm n}\rho_{\rm p}&\left[\frac{1}{2}(\mathbf{v}_{\rm n}^2-\mathbf{v}_{\rm p}^2)\right.\\&\left.+3\left(\frac{P_{\rm n}}{\rho_{\rm n}}-\frac{1}{2}\frac{P_{\rm p}}{\rho_{\rm p}}\right)\right],\nonumber \\
e_{\rm n}=&\frac{P_{\rm n}}{\gamma-1}+\frac{1}{2}\rho_{\rm n} v_{\rm n}^2.
\end{align}
The full nondimensional equations solved for the evolution of the charge-neutral ion-electron plasma fluid are:
\begin{align}
\frac{\partial\rho_{\rm p}}{\partial t}&+\nabla\cdot(\rho_{\rm p}\mathbf{v}_{\rm p})=0, \\
\frac{\partial}{\partial t}(\rho_{\rm p}\mathbf{v}_{\rm p})+\nabla\cdot&\left(\rho_{\rm p}\mathbf{v}_{\rm p}\mathbf{v}_{\rm p}+P_{\rm p}\mathbf{I}-{\mathbf{BB}}+\frac{\mathbf{B}^2}{2}\mathbf{I}\right)=\\
&\;\;\;\;\;\;\;\;\;\;\;\;\;\;\;\;\;\;\;\;\;\;\;\;\;\;\;\;\;\;\alpha_{\rm c}\rho_{\rm n}\rho_{\rm p}(\mathbf{v}_{\rm n}-\mathbf{v}_{\rm p})\nonumber,\\
\frac{\partial}{\partial t}\left( e_{\rm p}+\frac{B^2}{8\pi} \right)&+ \nabla\cdot\left[\mathbf{v}_{\rm p}(e_{\rm p}+P_{\rm p})-(\mathbf{v}_{\rm p}\times \mathbf{B})\times\mathbf{B}\right]=\\
+\alpha_{\rm c}&\rho_{\rm n}\rho_{\rm p}\left[\frac{1}{2}(\mathbf{v}_{\rm n}^2-\mathbf{v}_{\rm p}^2)+3\left(\frac{P_{\rm n}}{\rho_{\rm n}}-\frac{1}{2}\frac{P_{\rm p}}{\rho_{\rm p}}\right)\right]\nonumber,\\
\frac{\partial \mathbf{B}}{\partial t}&-\nabla \times (\mathbf{v}_{\rm p}\times \mathbf{B})=0\label{ind_eqn},\\
e_{\rm p}&=\frac{P_{\rm p}}{\gamma-1}+\frac{1}{2}\rho_{\rm p} v_{\rm p}^2,\\
\nabla\cdot\mathbf{B}&= 0\label{DB}.
\end{align}
Here the subscripts ${\rm n}$ and ${\rm p}$ differentiate between the neutral fluid and the plasma respectively. In this formulation $\alpha_{\rm c}\rho_{\rm a}=\nu_{\rm ab}$ which is the collision frequency of species $a$ on species $b$.
This system of equations has been non-dimensionalised in the following way: The velocity $\mathbf{v}$ has been non-dimensionalised using the sound speed of the combined fluids $C_{\rm s}$, the density $\rho$ by a reference total density $\rho_0$, and time $t$ by the inverse of a characteristic collision frequency $(\alpha_{\rm c, REF}\rho_0)^{-1}$. This implies that the lengthscales of the system are nondimensionalised by $C_{\rm s}(\alpha_{\rm c, REF}\rho_0)^{-1}$, the pressure $p$ by $C_{\rm s}^2\rho_0$ and the magnetic field $\mathbf{B}$ by $B_0/\sqrt{4\pi}=C_{\rm s}\sqrt{\rho_0}$. Here the subscript $0$ is used to represent a reference value of a quantity, $\gamma$ is the adiabatic index and $\beta$ is plasma $\beta$ (the ratio of gas to magnetic pressure calculated using the total gas pressure of the fluids).
We assume that both fluids follow the ideal gas law which in non-dimensional form become $T_{\rm n}=\frac{P_{\rm n}\gamma}{\rho_{\rm n}}$ and $T_{\rm p}=\frac{P_{\rm p}\gamma}{2\rho_{\rm p}}$, respectively.

The simulations are run using a fourth-order central difference method with the (P\underbar{I}P) code \citep{HILL2016}.
To allow the largest possible inertial range of any turbulent behaviour to develop, we do not include explicit viscosity or magnetic resistivity.
However, for stability of the scheme we employ an artificial viscosity and diffusion \citep{REM2009}.

\subsection{simulation setup}\label{setup}

In this set of simulations, we take the initial steady-state to be:
\begin{align}
\rho&=\begin{cases} 1 & \mbox{if $y < 0$};\\ 1.5 & \mbox{if $y > 0$}\end{cases},\\
v_x&=\begin{cases} -\Delta V \times1.5/2.5  & \mbox{if $y < 0$};\\ \Delta V \times1/2.5 & \mbox{if $y > 0$}\end{cases}, \; v_y=0,\\
p & =\frac{1}{\gamma},\\
B_x&=\sqrt{\frac{2}{\gamma\beta}}, \; \; \; B_y=0.
\end{align}
We use a velocity difference of $\Delta V=0.2$ and take plasma $\beta=200$ and $\gamma=5/3$.
We use an ionization fraction of $\xi_i= 0.015$ and a collisional coefficient of $\alpha_c= 300$.
The velocity magnitudes we have chosen here approximately put the linear instability into its zero momentum frame.

The system is perturbed by a white noise perturbation in the y component of the velocity field. The magnitude of the perturbation is limited to 0.1 per cent of the sound speed in the low density region.

We use a domain between $x=-1.5$ to $1.5$ and $y=-0.75$ to $0.75$ unless otherwise stated. 
For the high resolution simulations a grid of $16384\times 8192$ is used.
For all other simulations the grid is $2048\times 1024$. We use a periodic boundary for the $x$ boundaries, and symmetric boundaries that the magnetic field cannot penetrate for the $y$ boundaries.

\subsection{Physical meaning of this setup}\label{phys_mean}

{The non-dimesionalised growth rate $\sigma$ of the single-fluid ideal MHD instability (and as such is holds for our system in the limit $\alpha_{\rm c}\rightarrow\infty$) in the incompressible limit is given as:
\begin{equation}
    {\sigma^2}=k^2\frac{\rho_1}{\rho_1+\rho_2}\left(\frac{\rho_2}{\rho_1+\rho_2}\Delta V^2-\frac{4}{\gamma\beta\rho_1} \right),
\end{equation}
where $\rho_{1,2}$ are the densities below and above $y=0$ respectively, $k$ is the wavenumber of the perturbation, and $\Delta V$ is the velocity difference.
Note that this equation is equivalent to Equation 204 in Section 106 of \citet{CHAN1961} when gravity is set to zero, and as the initial conditions of this study have been put in the zero-momentum rest frame relating to the linear analysis, the growth rate $\sigma$ is either purely real or purely imaginary.
In the hydrodynamic limit, $\beta\rightarrow \infty$, this reduces to the purely hydrodynamic growth rate.}

For the parameters of the simulation we have $\sigma^2/k^2=0.0048$, which is half of the value for the purely hydrodynamic instability.
Though we are running a compressible simulation, the level of the compressibility (which can be estimated from the Mach number squared) is approximately $0.048$, so we expect that the estimated growth rates in the incompressible limit will reasonably hold.
However, when $\sigma^2$ is calculated purely for the plasma, the term associated with the magnetic tension is $\xi_i^{-1}$ larger.
In this case $\sigma^2/k^2=-0.31$, so though the neutral fluid is unstable and the coupled fluids are unstable (though less so than the neutral fluid), the plasma is stable. 
This means that the neutral fluid will always be unstable, but only when the perturbation is on a large enough scale the instability will involve the the whole fluid and the magnetic field.

{As we are looking at the nonlinear problem, it is also important to estimate the relative importance of the magnetic field in the nonlinear stage of the instability.
To do this, we follow methods that have been applied to the magnetic Rayleigh-Taylor instability\citep{HILL2016b}.
Firstly, we look at the saturation by secondary KHi.
For the displacement of the boundary $\eta_y$ we can estimate, using the linear eigenfunction \citep[e.g.][]{CHAN1961}: 
\begin{equation}
    \eta_y(\mathbf{x},t)=\eta_y(0)\exp(ikx-k|y|+\sigma t),
\end{equation}
the displacement at which the magnitude of the terms that are second-order in $\eta_y$ are of the same order of magnitude as the first-order terms.
Firstly, we look at how the development of secondary shear-flows can saturate the instability:
\begin{equation}
    \frac{\partial^2 \eta_y}{\partial t^2}=\sigma^2 \eta_y\approx\sigma^2\eta_y\frac{\partial \eta_y}{\partial y}=\sigma^2k\eta_y^2.
\end{equation}
This gives:
\begin{equation}
    \eta_y\approx\frac{1}{k},
\end{equation}
which is the same as the characteristic vertical scale given in the linear eigenfunction.
If we take the Lorentz force as the nonlinear term then:
\begin{equation}
    \frac{\partial^2 \eta_y}{\partial t^2}=\sigma^2\eta_y\approx \frac{1}{\rho}(\mathbf{j}\times\mathbf{b})_{\rm y}\approx\frac{2}{\rho_1+\rho_2}kb_{y}^2,
\end{equation}
where the $\mathbf{j}$ and the $\mathbf{b}$ are lower case to denote perturbed components of the current and magnetic field respectively, the subscript $y$ on the vector field to show which component of the vector is used, and the density $\rho$ has been taken to be $(\rho_1+\rho_2)/2$.
In linear MHD \citep{HILL2019}
\begin{equation}
    b_y=\frac{ik}{\sigma}B_yv_y=ik B_x\eta_y.
\end{equation}
Therefore,
\begin{equation}
    \sigma^2\eta_y
        \approx\frac{2\rho_1}{\rho_1+\rho_2}V_A^2k^3\eta_y^2,
\end{equation}
where $V_A=B_x/\sqrt{\rho_1}$.
This gives (for the initial conditions of this study):
\begin{equation}
    \eta_y\approx\frac{\rho_1+\rho_2}{\rho_1}\frac{\sigma^2}{2V_A^2k^3}=\frac{\gamma\beta\rho_1}{4k}\left(\frac{\rho_2}{\rho_1+\rho_2}\Delta V^2-\frac{4}{\gamma\beta\rho_1}\right)=\frac{1}{k},
\end{equation}
which is the same length as found for hydrodynamic nonlinearities.}
Therefore, we expect that for an MHD system with the initial parameters of our model, the magnetic nonlinearities will be of the same level of importance as flow nonlinearities in determining the dynamic evolution of the system.
This is also presented through the reference numerical simulations in Section \ref{ref_solns}.

To estimate the spatial scales at which decoupling will occur in these calculations for both linear and nonlinear process, we can perform a comparison of timescales.
Firstly, the characteristic collision frequencies between the species are $\nu_{pn}=300\times0.985=295.5$ and $\nu_{np}=300\times0.015=4.5$ based on the initial values of $\xi_{\rm i}=0.0015$ and $\alpha_{\rm c}=300$.
Comparing the collision frequency of neutrals onto ions with the growth rate of the hydrodynamic KHi we find a non-dimensional wavenumber of $k\approx46$ where they are equal.
Comparing the collision frequency of ions onto neutrals with the magnitude ion surface Alfv\'{e}n wave frequency we find a wavenumber of $k\approx531$ where they are equal.
For wavenumbers smaller than these we would expect to see more coupled, single fluid like behaviour, and greater than these would behave in a more decoupled fashion (note that the $k$ used in the high resolution simulation is between $\sim2.1$ to $\sim17158$).

For nonlinear phenomena, we can look to arguments of turbulence scalings.
In terms of having a turbulent cascade driving a vortex that is in the neutral fluid and uninhibited by the magnetic field, then the frequency of the turbulence given by:
\begin{equation}
    \nu_{\rm turb}\propto(\epsilon k^2)^{1/3},
\end{equation}
where $\epsilon$ is the energy cascade rate, would have to be greater than $\nu_{ni}=4.5$.

To summarise the information of this subsection, given to help the reader understand why the particular parameter regime is under study, we have chosen this particular parameter regime because we expect it will allow us to look at the coupling from strong coupling at the largest spatial scales, to completely decoupled at the smallest.
This parameter regime also has the benefit of having an important contribution by the magnetic field in determining both the linear and nonlinear behaviour of the system.

\section{Results}

\subsection{Reference HD and MHD simulations}\label{ref_solns}

To provide sufficient context to our solutions we have performed both HD and MHD simulations of the initial conditions laid out in Section \ref{setup} for a resolution of $2048\times1024$ over a domain of $x=[-3,3]$ and $y=[-0.75,0.75]$. These are performed to provide context for the two-fluid simulations, showing the evolution we expect in single fluid simulations.

\subsubsection{HD evolution}

\begin{figure*}
  \centerline{
    \includegraphics[width=17cm]{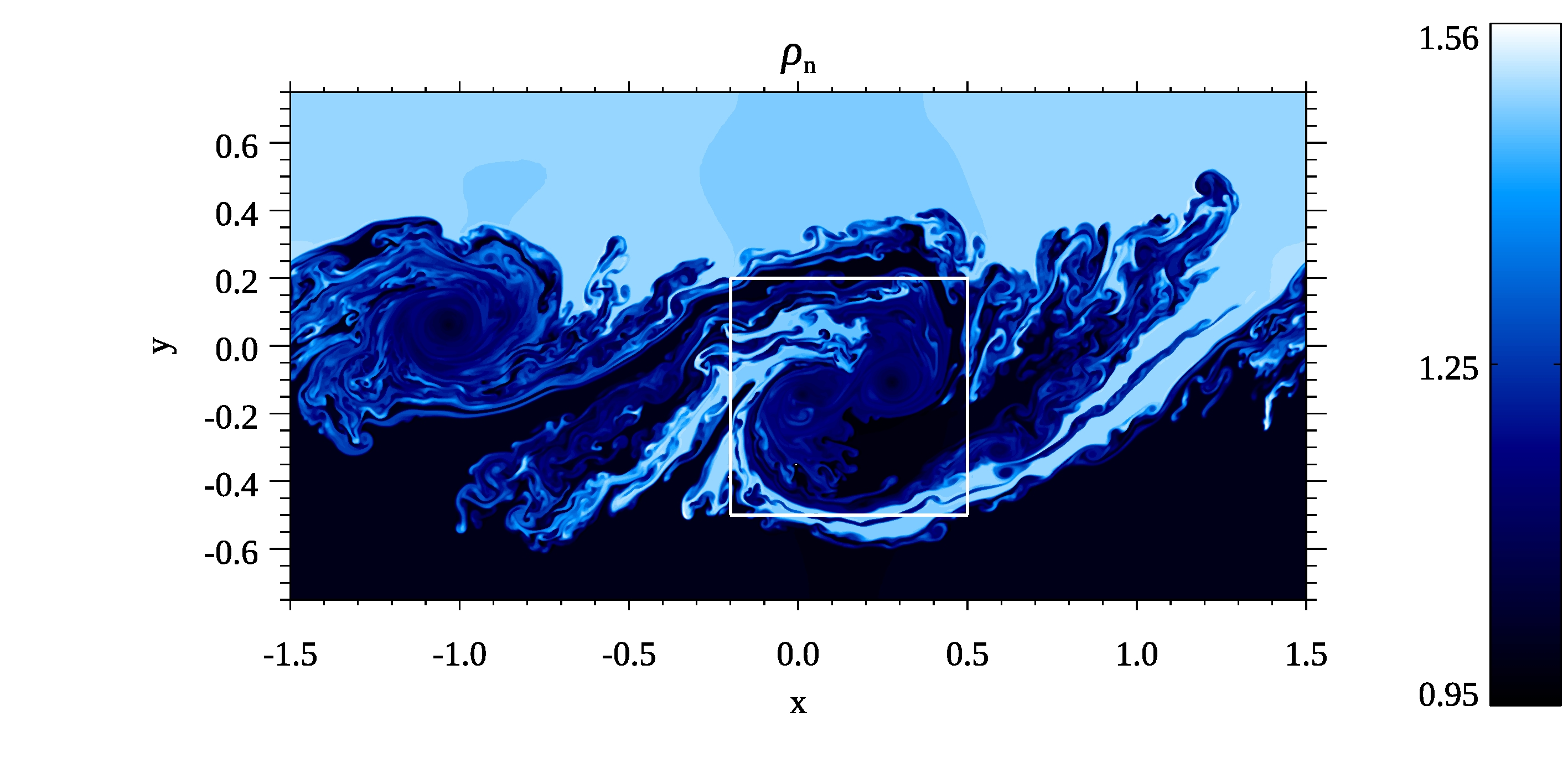}}
  \caption{$\rho_{\rm n}$ distribution {at $t=50$} showing the nonlinear stage of the HD KHi for the chosen parameters. Both large scale vortices ($\sim$ half the box width) and smaller turbulent structures are visible.}
\label{HD_fig}
\end{figure*}

In the HD case, the nonlinear evolution is characterised by the formation and subsequent merger of vortices. 
Figure \ref{HD_fig} shows the nonlinear development of the simulation at time $t=50$.
There are a collection of vortices that have formed, and at $x\approx 0.2$ and $y\approx -0.1$ there is an example of two vortices undergoing a merger (see the white box on Figure \ref{HD_fig}).
As a result of the density contrast, it is clear that as well as the inverse cascade creating the larger vortices, there is also a cascade of scales driving the formation of many small vortices.

\begin{figure}
  \centerline{
    \includegraphics[width=9cm]{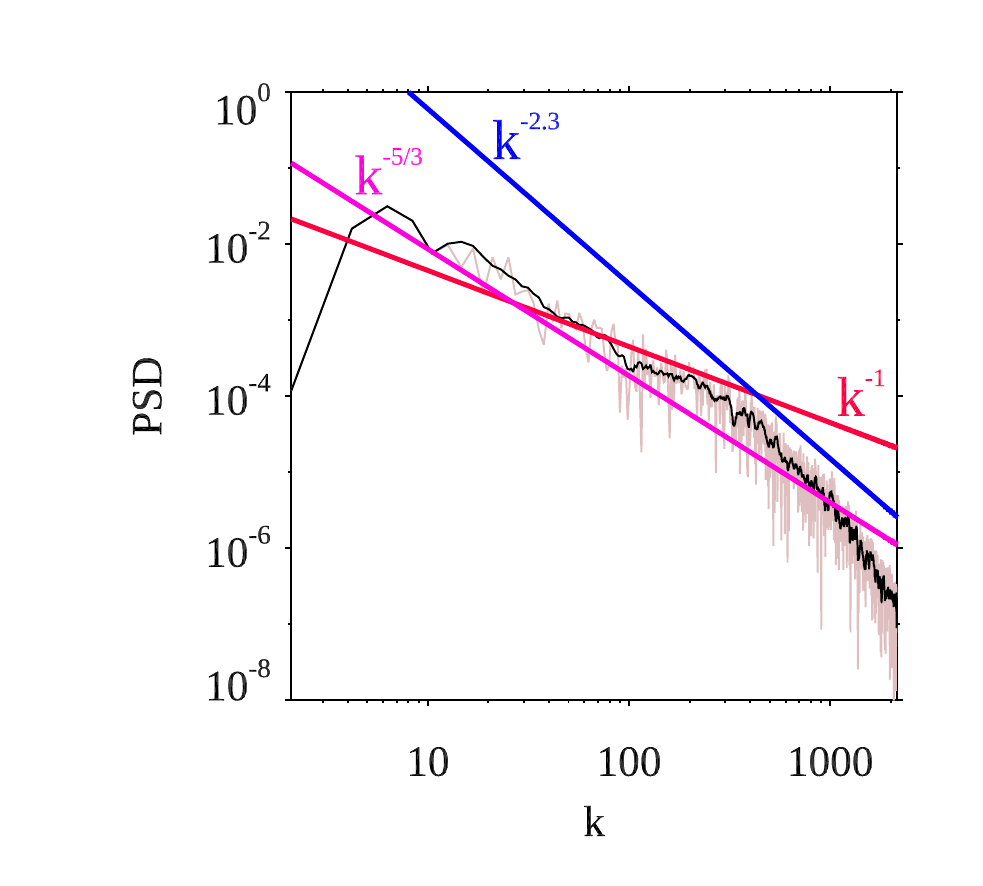}}
  \caption{Power spectra for the density weighted y velocity at $y=0$ {and $t=50$}.     
  The black line showing the smoothed spectra with the unsmoothed spectra plotted behind it. The straight lines show the slopes of $k$ to different exponents used to highlight regions of different behaviour in different regions of the spectra.}
\label{HD_spectra_fig}
\end{figure}

A snapshot of the density weighted velocity ($\sqrt{\rho} v_y$) spectra is shown in Figure \ref{HD_spectra_fig}. 
At larger $k$ the spectra has a relatively constant slope over a region of about a decade (between $k\approx 4$ to  $k\approx 100$), this approximately corresponds to a slope of $\sim k^{-5/3}$ as shown by the magenta line on the figure.
At smaller scales a bump in the spectra is present, composed of a region which follows a slope of $\sim k^{-1}$ at smaller $k$ and $\sim k^{-2.3}$ at larger $k$.

Previous studies of the spectra of 2D hydrodynamic KHi mixing\citep{MATSU2004}, provide an important reference point for the slopes present in our spectra.
For similar density contrasts to the one used in this paper, spectra containing a dominant component that followed $k^{-1.38}$, which is consistent with the approximate relation with $k^{-5/3}$, that we found. This approximately Kolmogorov-like spectra was formed as a result of the cascade driven by the baroclinic term in the vorticity equation ($\propto \nabla\rho\times\nabla p$). 

Looking at the results shown in Figure 2 of \citet{MATSU2004}, as their system is evolving in time towards the statistical steady state, there are bumps present in the spectra. We conjecture that the departures from this slope we see (in the form the approximate slopes of $\propto k^{-1}$ and $\propto k ^{-2.3}$ components), are present because the system is still undergoing an inverse cascade, and as such has not reached a statistical steady state. Evidence for this conjecture can be seen in Figure \ref{HD_fig} where there are two vortices in the process of merger. 

\subsubsection{MHD evolution}

\begin{figure*}
  \centerline{
    \includegraphics[width=17cm]{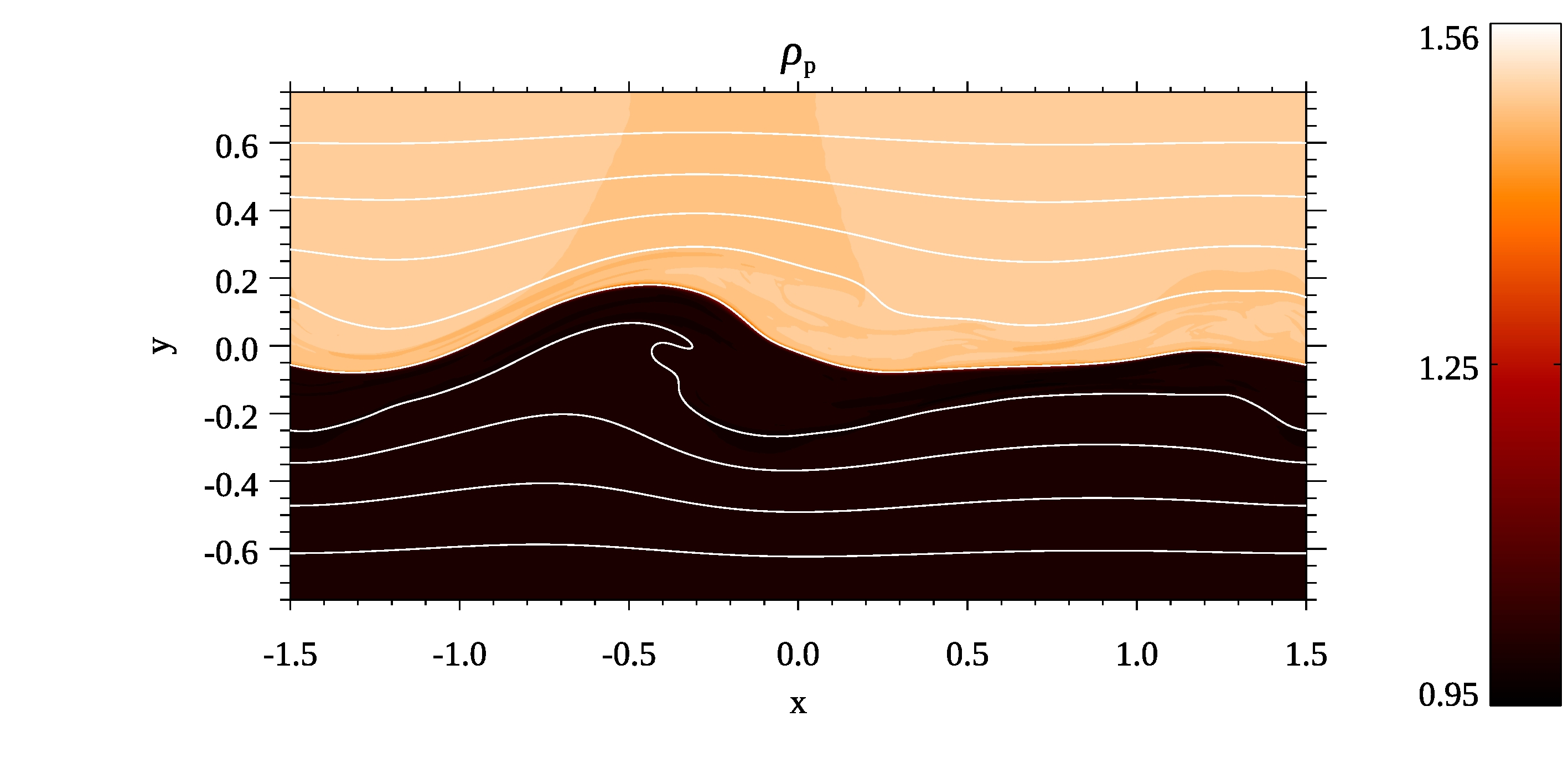}}
  \caption{$\rho_{\rm p}$ distribution {at $t=50$} showing the nonlinear stage of the MHD KHi for the chosen parameters. White lines show the magnetic field lines. Note that no vorticies have formed, but there is some evidence of the magnetic field wrapping up.}
\label{MHD_fig}
\end{figure*}

The nonlinear stages of this instability, for the parameters we have chosen, shows a marked difference from the purely HD simulation.
Figure \ref{MHD_fig} shows the density distribution of the simulation at the same time as shown in Figure \ref{HD_fig}.
From the large undulations of the boundary between the different density layers, it is clear that there has been growth of a linear instability.
However, there has not been the development of over-turning vortices that the KHi is associated with (and are visible in the HD simulation).
The key reason behind this is that the magnetic field is playing a key role in the nonlinear suppression of the instability.
Therefore we can understand that the MHD case is linearly unstable, but in a sense nonlinearly stable. This situation has been seen in previous simulations \citep{RYU2000}.

It is not that the dynamics are purely limited to some large deformations of the boundary. The magnetic field lines, shown in the figure by the white lines, trace the smooth boundary between the two fluids accurately. However, in the regions just above and below the interface, the magnetic field has been more significantly disturbed.
We can understand this by thinking about how the flow is trying to behave, and how the magnetic field is working against this.
If we imagine the fluid flowing around a hump of the interface, the fluid needs to experience a force to make it keep following the curve of the interface. As the gas pressure is working to form vortices, it is magnetic tension that must provide the force to work against the gas pressure. Therefore, the magnetic field undergoes a distortion preventing the formation of vortices.

\begin{figure*}
  \centerline{
    \includegraphics[width=9cm]{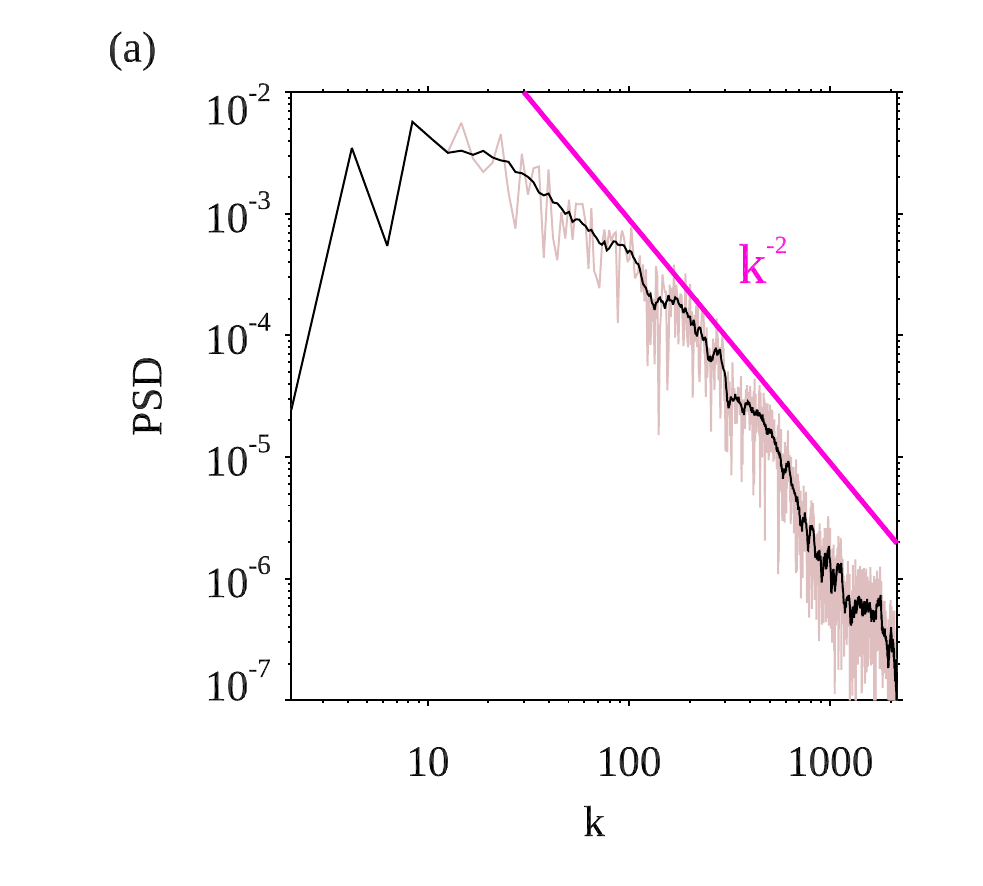}
    \includegraphics[width=9cm]{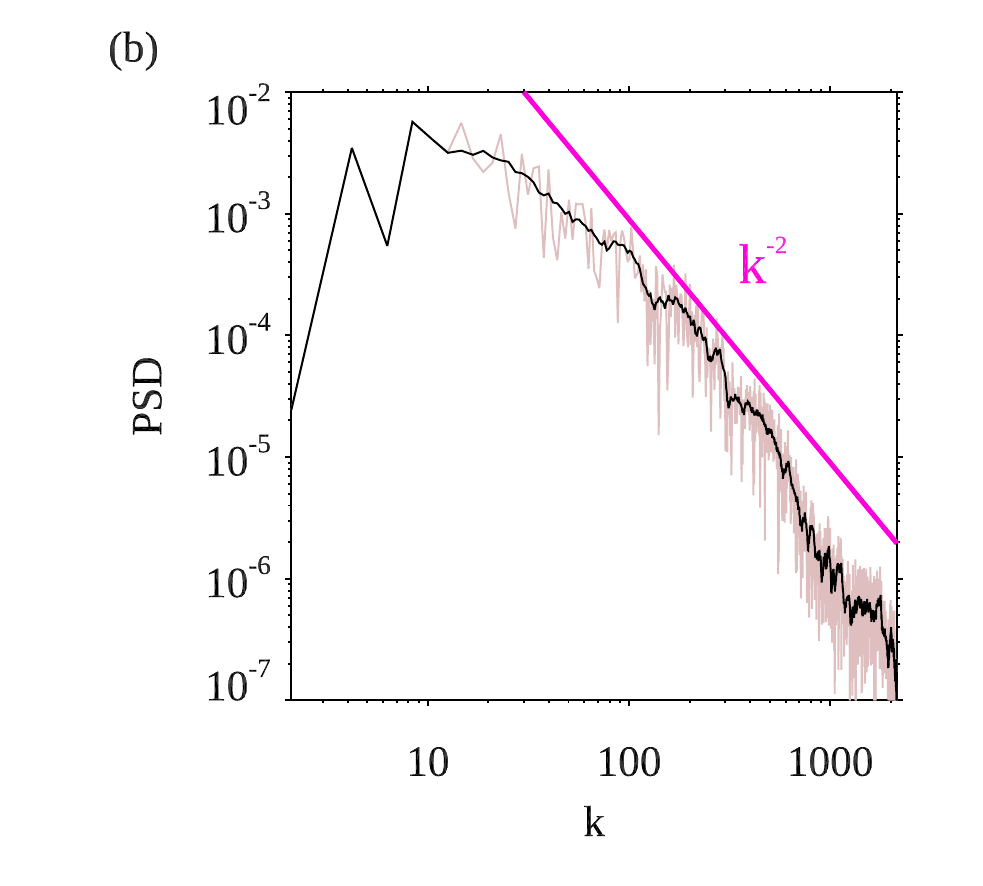}}
  \caption{Power spectra for the density weighted neutral y velocity (panel a) and the y component of the magnetic field (panel b) at $y=0$ {and $t=50$}.     
  The black line showing the smoothed spectra with the unsmoothed spectra plotted behind it. The straight line show the slopes of $k$ to the exponent $-2$ used to highlight the behaviour of the spectra.}
\label{MHD_spectra_fig}
\end{figure*}

Figure \ref{MHD_spectra_fig} displays the density weighted velocity spectra $\sqrt{\rho}v_y$ and $B_y$ at $y=0$.
As with the reference HD case the smoothed spectra is shown in black with the unsmoothed spectra behind it. The peak magnitude of the spectra in the MHD case are about one order of magnitude smaller than the HD case.
The spectra can be characterised by a flat region at small $k$, which falls away beyond $k>20$.
The magenta line shows the slope expect for a spectra that follows $k^{-2}$. The central region of the power spectra between $k$ of $100$ and $500$ appears to roughly follow this exponent. However, in this case we could have chosen different regions and found any slope with exponent between $-1$ and $-2.5$.

\subsection{High-Resolution PIP Simulations}

Next we look at the nonlinear dynamics of the nonlinear KHi in a partially ionised plasma at multiple coupling scales and the cross-scale coupling, through the evolution of a high-resolution simulation.
Using the dynamics present in the reference HD and MHD cases, we can expect that in cases where the neutral fluid has decoupled from the magnetic field it would be expected to have the full nonlinear development of the instability in the neutral fluid. However, in situations where the coupling is stronger, then the fully nonlinear instability will not develop.
By studying the dynamics over multiple lengthscales then the role of this coupling (which naturally becomes stronger at larger scales) can be elucidated.

\begin{figure*}
  \centerline{
    \includegraphics[width=17cm]{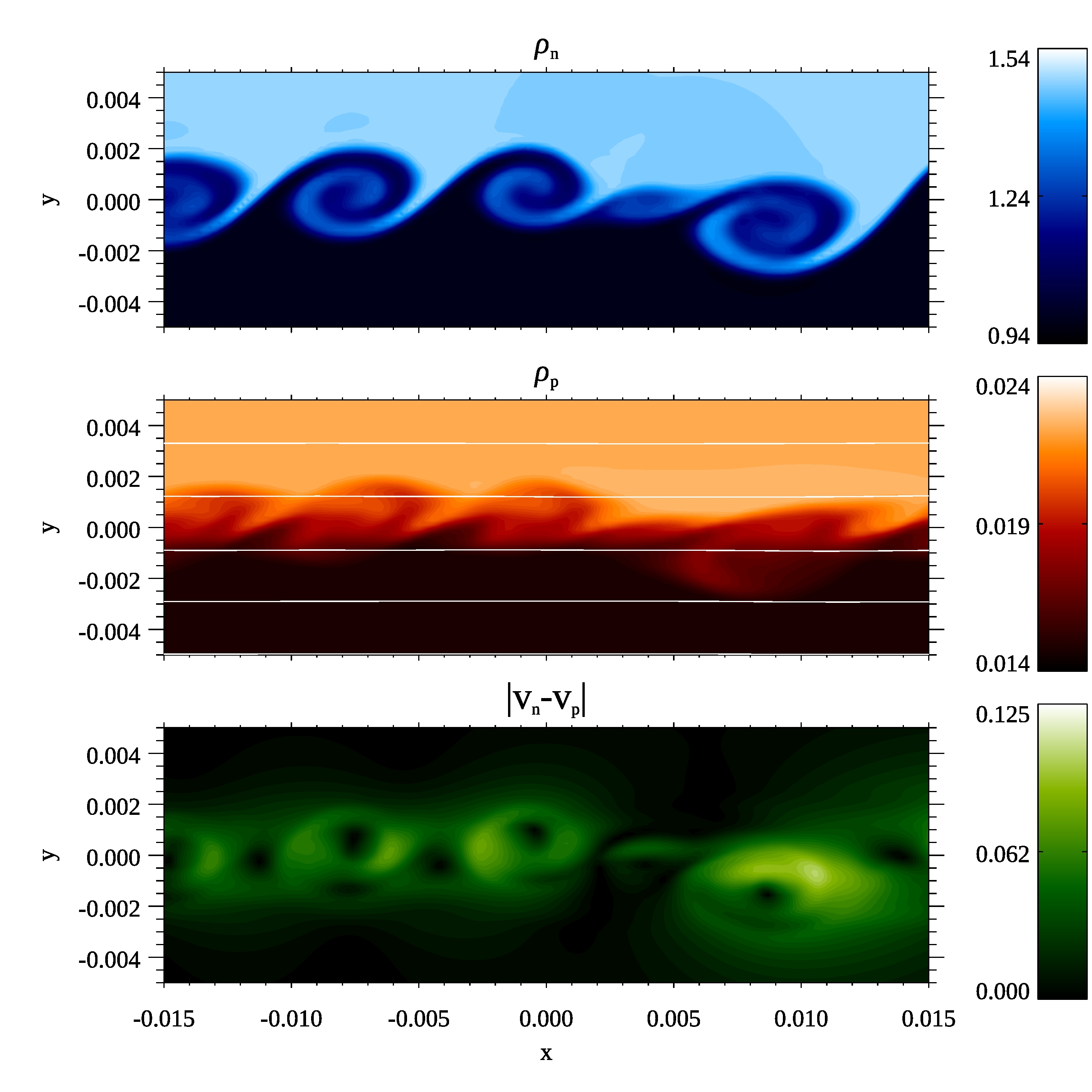}}
  \caption{Zoomed in snapshot of the simulated densities (top panel for $\rho_{\rm n}$ and middle panel for $\rho_{\rm p}$) and magnitude of the drift velocity (bottom panel) at $t=0.4$. The white lines in the middle panel show the magnetic field lines. A number of vorticies are clear in the neutral density, but there is only a diffuse plasma density and no response in the magnetic field. Velocity differences of magnitude $\sim \Delta V/2$ are created as a result of the decoupling of the neutral fluid.}
\label{t1_fig}
\end{figure*}

Figure \ref{t1_fig} shows the system at $t=0.4$ at scales where $\nu_D\sim\nu_{\rm pn}>\nu_{\rm np}$ (where $\nu_D$ is the dynamic frequency). Here we have zoomed into the region of $x=[-0.015,0.015]$ and $y=[-0.005,0.005]$.
This early stage is characterised by the quick development of the instability in the neutral fluid at very small scales. 
Though there is some response in the plasma, but on measuring the magnitude of $B_y/B_x$ at $y=0$ is bounded by $|B_y/B_x|\lesssim 0.01$ showing that the magnetic field has not been significantly deformed. Therefore, at these scales we are seeing an almost purely hydrodynamic evolution which will mimic that seen in the HD reference case.

The result of the different dynamics in the two fluids results in substantial drift velocities in the system.
The magnitude of the drift in the velocity between the neutral and plasma fluids reaches values up to approximately 60 per cent of the initial velocity shear.
The largest drift velocity magnitudes appear where the neutral fluid flow is roughly perpendicular to its original flow direction, i.e. where the neutral fluid contained in a vortex is moving directly across the magnetic field. 

\begin{figure*}
  \centerline{
    \includegraphics[width=17cm]{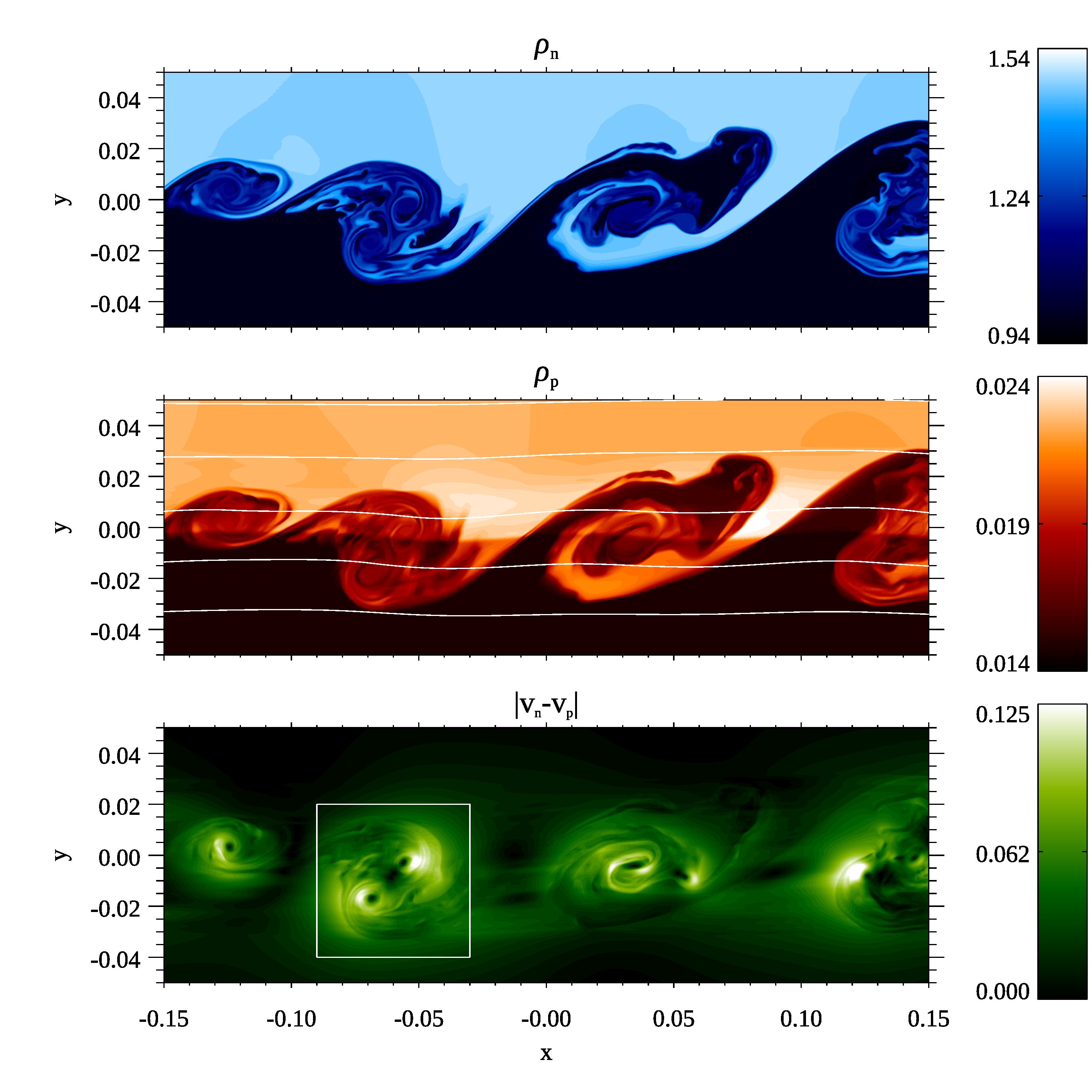}}
  \caption{Zoomed in snapshot of the simulated densities (top panel for $\rho_{\rm n}$ and middle panel for $\rho_{\rm p}$) and magnitude of the drift velocity (bottom panel) at $t=2.25$. The white lines in the middle panel show the magnetic field lines. A number of vortices of multiple scales are clear in the neutral density. These structures are also clear in the plasma density even though there has been no response in the magnetic field. Velocity differences of magnitude $\sim \Delta V/2$ are created as a result of the decoupling of the neutral fluid.}
\label{t2_fig}
\end{figure*}

Figure \ref{t2_fig} shows the system at $t=2.25$ at scales where $\nu_{\rm pn}>\nu_{\rm D}>\nu_{\rm np}$.
Here we have zoomed into the region of $x=[-0.15,0.15]$ and $y=[-0.05,0.05]$.
At these scales there is still fully-developed vortex formation in the neutral fluid.
It also appears that the plasma fluid is responding to the neutral fluid motions, with vortex-like density structures forming in the plasma density.
However, looking at the magnetic fieldline distribution, even though the vortex-like density structures have formed the magnitude by which the magnetic field has been distorted (the magnitude of $B_y/B_x$ at $y=0$ is $|B_y/B_x|< 0.1$) appears to be significantly smaller than that of the neutral density interface. 
This gives the impression that the plasma is flowing directly across the magnetic
field. This is analysed in Section \ref{plasma_flow}.

At these scales, there are still large regions with significant velocity drift. These regions reach a maximum value of approximately 80 per cent of $\Delta V$. This is strongly localised in the vortices to regions near the core of the vortex (as defined by the scale of the density structure).
The two merging vortices at $[-0.06,-0.01]$ (see region in white box in the bottom panel of Figure \ref{t2_fig} highlighting the drift velocity structure associated with vortex merger) show the interesting feature of almost no velocity drift between them, but on the outer edges of the vortices there is large velocity drift. 

\begin{figure*}
  \centerline{
    \includegraphics[width=17cm]{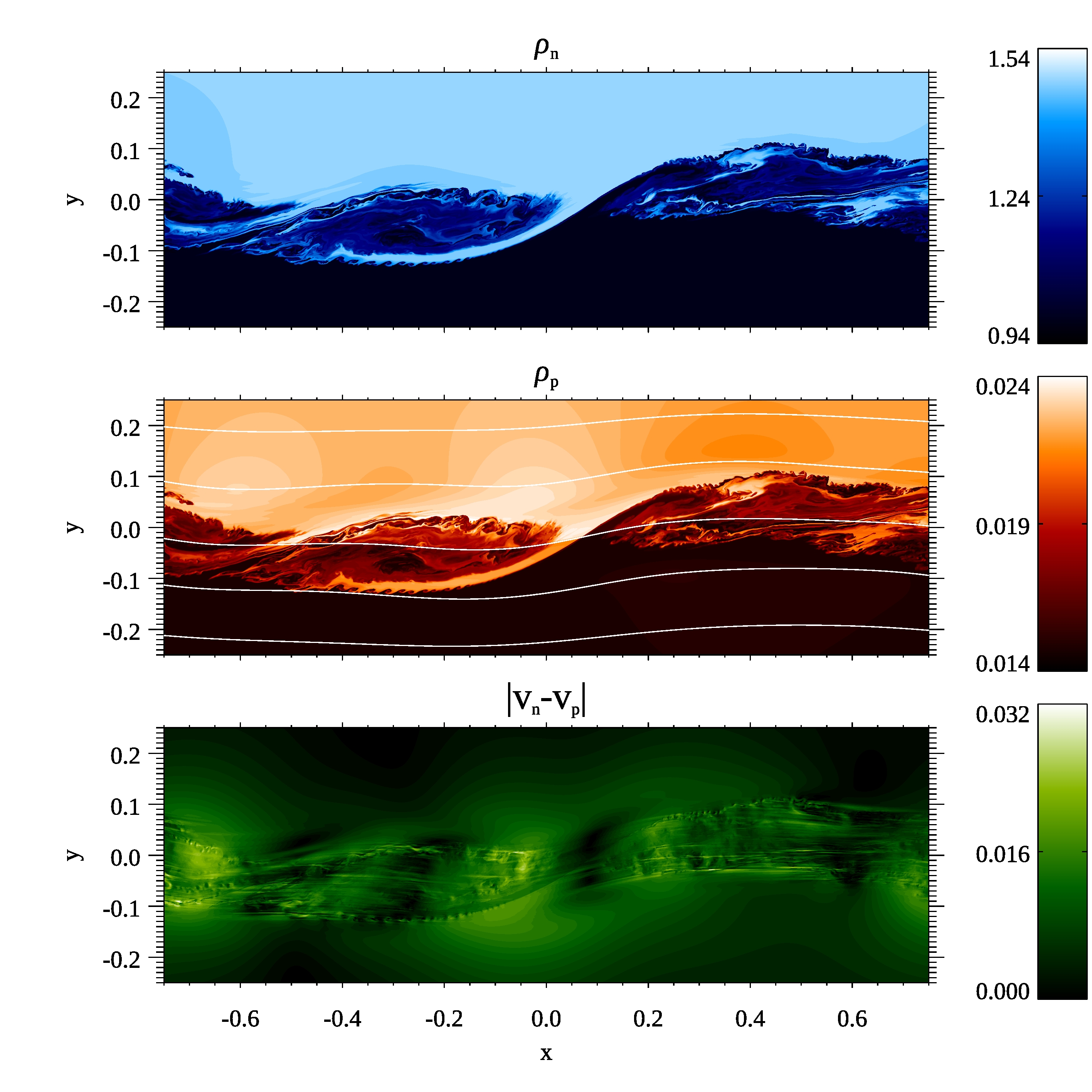}}
  \caption{Zoomed in snapshot of the simulated densities (top panel for $\rho_{\rm n}$ and middle panel for $\rho_{\rm p}$) and magnitude of the drift velocity (bottom panel) at $t=19$. The white lines in the middle panel show the magnetic field lines. Multiple scale vortices are clear in both the neutral density and in the plasma density. There is a large scale response in the magnetic field, but nothing connected to the other smaller scales present. Smaller velocity difference magnitudes of $\sim \Delta V/5$ are created as a result of the decoupling of the neutral fluid.}
\label{t3_fig}
\end{figure*}

Figure \ref{t3_fig} shows the system at $t=19$ at scales where we estimate that $\nu_{\rm pn}>\nu_{\rm D}\sim\nu_{\rm np}$.
Here we have zoomed into the region of $x=[-0.75,0.75]$ and $y=[-0.25,0.25]$.
There are fully-formed vortices in the neutral fluid, with many smaller secondary vortices formed.
At these scales the influence of the magnetic field is beginning to be felt by the neutral fluid as a result of the collisions between the fluids. 
This can be observed in the shape of the largest vortices that have formed.
The large scale vortices are squashed and elongated along the magnetic field,
but the small-scale secondary instabilities appear unaffected by the magnetic field.
The structure of the vortices, including the secondary vortices, is also clearly visible in the plasma density distribution.
Note that this snapshot is taken from the same simulation as used for Figures \ref{t1_fig} and \ref{t2_fig}, just taken at a later time and zoomed out to display larger dynamics.

At this larger scale, the absolute magnitude of the drift velocity has decreased (note the reduction in range of the colourscale). The general structure shows the vorticies and the disturbances they have made outside of the instability layer. 
There is also noticeable small-scale structuring in the drift velocity created by the secondary instabilities, but the large scale structure, as well as highlighting the vortex shape, contains contains long, coherent strips that follow the direction of the magnetic field.

\subsubsection{How can the plasma move across the magnetic field?}\label{plasma_flow}

One question that can be raised about these results is: how does the density structure form in the plasma if the magnetic field is not bending? This may lead to the impression that the plasma is moving across the magnetic field, which should not be possible when the induction equation contains no terms that allow a decoupling of the plasma and the field, i.e. there is no magnetic diffusion.

\begin{figure*}
  \centerline{
    \includegraphics[width=19cm]{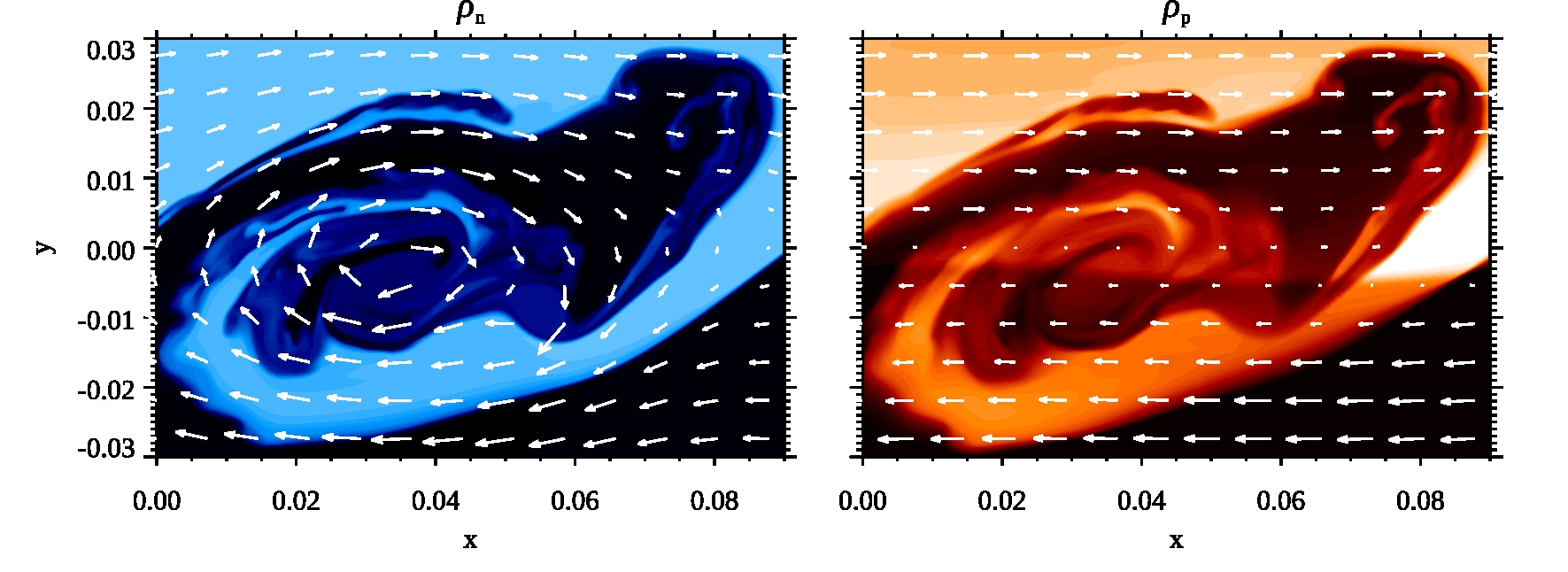}}
  \caption{Image of the neutral and plasma density structures for a vortex isolated from Figure \ref{t2_fig} {(i.e. taken at $t=2.25$}). The white arrows show the neutral and plasma velocity vectors respectively. The neutral flow shows a clear vortex, but the plasma flow is still a shear flow. Note that the swirl in the neutral density maintains its level, but in the plasma density there is a jump in the magnitude of the density at $y\approx 0$.}
\label{slip_fig}
\end{figure*}

In short, the plasma is still coupled to the field. It is only the neutral fluid that moves across the magnetic field. 
This is shown in Figure \ref{slip_fig} where we zoom in to the range $x=[0,0.09]$ and $y=[-0.03,0.03]$ to analyse one vortex structure as shown in Figure \ref{t2_fig}.
The colour contour shows the density of the neutral fluid (left panel) and plasma (right panel) and the arrows show the velocity vector.
If we look at the velocity field of the neutral flow (from t=2.25), a clear vortex has formed along with the vortical density structure. However, in the plasma, the flow vector direction is still close to the original shear flow (i.e. aligned with the $x$-axis) and shows no vortex.
Therefore, even though the density shows a vortex, the flow is still frozen into the magnetic field. 

The question then is: why does this density structure exist? 
This all happens as a thermal effect. When the cool dense neutral fluid (initially set in the $z>0$ region) moves into the $z<0$ region (or the light warm neutral fluid moves to the $z>0$) they force the plasma they interact with to change temperature. The plasma pressure gradient along the magnetic field this creates drives compression or expansion of the plasma along the field lines and the cooling or heating of the instability region leads to contraction or expansion of the layer in the direction across the magnetic field. This leads to increases or decreases in the plasma density that match the high and low density patches of neutral fluid.

\begin{figure*}
  \centerline{
    \includegraphics[width=16cm]{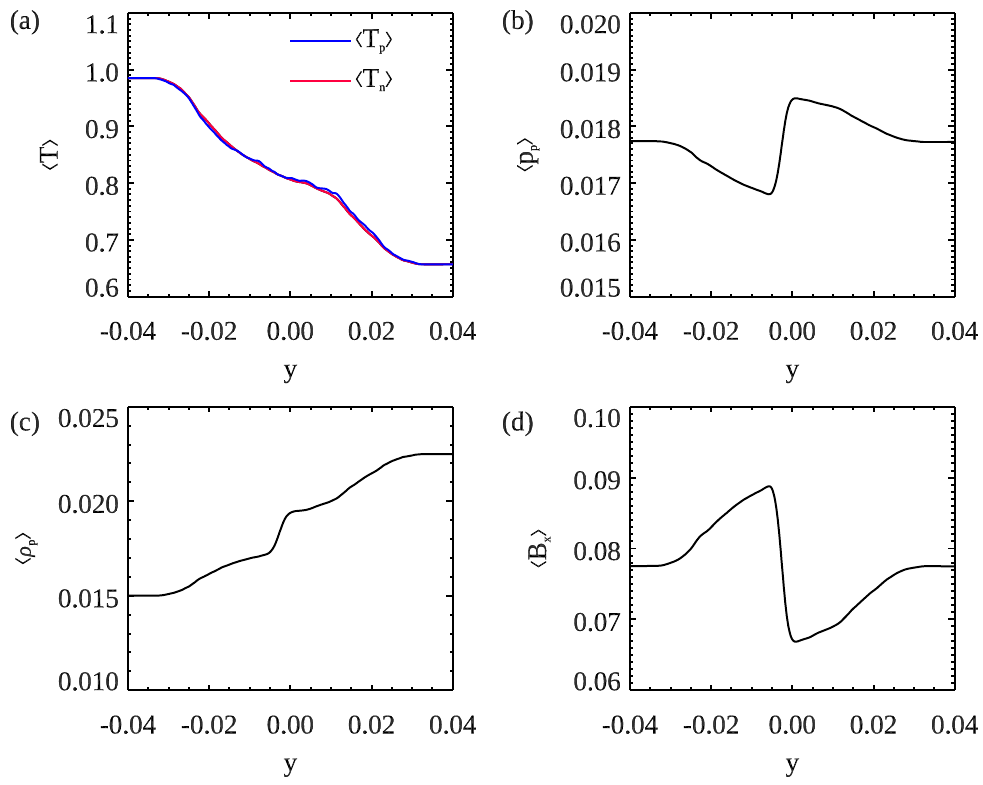}}
  \caption{Plots of the $x$ averaged temperature (panel a), plasma pressure (b), plasma density (c) and x magnetic field (d) {at t$2.25$}.}
\label{mean_pressure}
\end{figure*}

The proof of this is shown in Figure \ref{mean_pressure}, where the $x$-direction mean of the temperature ($T$), $P_{\rm p}$, $\rho_{\rm p}$ and $B_y$ are shown. The mean temperature of both fluids are plotted, but they follow the same distribution as a result of the heating/cooling effect the neutral dynamics have on the plasma. This results in regions of increased/decreased plasma pressure, which correspond to regions of decreased/increased plasma density and magnetic field strength as a result of the squeezing and expanding of the layer.

There is an interesting corollary that results from this: even if the motions in the neutral fluid are very close to being incompressible, the inclusion of the magnetic field results in the plasma fluid undergoing compressible motions.
Therefore, even only weakly compressible turbulent motions should be treated as compressible in a two-fluid system because of the response of the plasma to the neutral motions.

As this is a thermal effect, then we can expect that its magnitude will vary greatly with plasma $\beta$, especially the plasma $\beta$ when calculated with respect to the plasma and not the bulk fluid (here we denote this $\beta_{\rm p}$ which equals for our nondimensional variables $\beta_{\rm p}=2p_{\rm p}/B^2$). To understand the magnitude of this effect, we can make some simple estimates. Because the temperature of the plasma is determined by that of the neutral fluid, once the cooling or heating process as a result of the neutral fluid acting as a sink or source has occurred, the subsequent evolution can be treated as isothermal (i.e. the temperature of the plasma is fixed as that of the neutral fluid). Therefore the ratio of the mean pressure in a layer ($\langle p_{\rm p}\rangle_1$) against the pressure before contraction ($\langle p_{\rm p}\rangle_0$) will be $l_0/l_1$ where $l_0$ is the original layer width and $l_1$ the new width. Due to flux conservation this will also apply to the magnetic field giving $\langle B_x\rangle_1/\langle B_x\rangle_0=l_0/l_1$. If the temperature has dropped by a factor of $\delta_{\rm T}$, then $\langle p_{\rm p}\rangle=\delta_T p_{0,p}$, the contraction of the layer will be given by the following total pressure in the layer
\begin{align}
    \frac{\langle B\rangle_1^2}{2}+\langle p_{\rm p}\rangle_1&=\frac{\langle B\rangle_0^2}{2}\left(\frac{l_0}{l_1}\right)^2+\langle p_{\rm p}\rangle_0\frac{l_0}{l_1}\nonumber\\&=\frac{\langle B\rangle_0^2}{2}\left(\frac{l_0}{l_1}\right)^2+\delta_{\rm T} p_{0,p}\frac{l_0}{l_1}.
\end{align}
Balancing this with the external pressure leads to
\begin{equation}
    \frac{\langle B\rangle_0^2}{2}\left(\left(\frac{l_0}{l_1}\right)^2+\delta_{\rm T} \beta_{\rm p}\frac{l_0}{l_1}\right)=\frac{\langle B\rangle_0^2}{2}(1+\beta_{\rm p}).
\end{equation}
Therefore the change in density (given by conservation of mass) is given by
\begin{equation}
    \frac{\langle \rho_{\rm p}\rangle_0}{\langle \rho_{\rm p}\rangle_1} =\frac{l_1}{l_0}=\frac{1}{2(1+\beta_{\rm p})}\left(\delta_{\rm T} \beta_{\rm p}+\sqrt{\delta_{\rm T}^2\beta_{\rm p}^2 + 4(1+\beta_{\rm p})}\right),
\end{equation}
where $\rho'$ is the new density.

In the $\beta_{\rm p}=0$ limit this ratio just becomes $1$, and in the $\beta_{\rm p}=\infty$ limit this ratio becomes $\delta_{\rm T}$, i.e. in low $\beta_{\rm p}$ situations, the global contraction of the layer will not occur. For the case in Figure \ref{mean_pressure} the mean temperature drops by 20 per cent in the $y<0$ region ($\delta_{\rm T}=0.8$), so our simple estimate for the mean density in this region (around $y=-0.01$) would be $\langle \rho_{\rm p}\rangle = 0.0175$ which is fairly accurate given the actual value is $\langle \rho_{\rm p}\rangle = 0.0170$.

This explains how the material squeezes in the $y$ direction, but the remaining question is how large do the fluctuations along a field line get from this new mean value? The key is to understand how much mass is transported into the local regions where the pressure is reduced as it cools to match the neutral temperature, or removed in the regions of higher temperature.
For example, to get an average temperature halfway between the initial temperatures on either side of the discontinuity (as is roughly shown in Figure \ref{mean_pressure}) means that half of the field line is filled with material that is cooler than it was originally. 
As the expansions and contractions are isothermal, due to the arguments in the previous paragraph, the pressure is predetermined. Also magnetic effects cannot be important as these dynamics occur purely along the magnetic field. 
After the expansions and contractions along the fieldline, the pressure along the fieldline will be uniform (given by $p'$) and the higher density region will occupy half the fieldline (a length of $l'$).
If the pressure changes are given by:
\begin{equation}
    \frac{p'}{p_1}=\frac{l_1}{l'}, \; \frac{p'}{p_2}=\frac{l_2}{l'}, 
\end{equation}
where $l_1+l_2=2l'$.
Here $p_1$ is the plasma pressure in the cooled region given by $\delta_2\langle p_{\rm p}\rangle_1$ where $\delta_2$ is given as the ratio of the temperature of the cooled/heated material to the background temperature, and $p_2=\langle p_{\rm p}\rangle_1$.
The force balance is given by
\begin{equation}
    \delta_2\frac{l_1}{l'}=\frac{l_2}{l'},
\end{equation}
which means the density change is
\begin{equation}
    \frac{\rho_{\rm p}'}{\langle \rho_{\rm p}\rangle_1 }=\frac{l_1}{l'}=\frac{1}{\delta_2+1}.
\end{equation}
For the example, for the case shown in Figure \ref{t2_fig}, where we know the mean density around $y=-0.01$ is $\langle \rho_{\rm p}\rangle_1\approx 0.0175$, as $\delta_2=2/3$ we predict $\rho_1'\approx 0.021$ in the neutral vortex region, which tallies with the values shown in the figure.
This argument implies that even when the plasma $\beta$ is close to zero, there will still be some response in the plasma density to the neutral dynamics even when the system is magnetically dominated.

\subsubsection{The coexistence of scales at different coupling levels}

In Figure \ref{t3_fig} it is clear that many smaller vortices have formed as a result of the presence of a cascade of energy. The energy cascade that creates the smaller scale vortices in this case is driven by the presence of gradients in the density allowing for a non-zero baroclinic term ($\propto \nabla \rho \times \nabla p$) \citep[e.g.][]{MATSU2004}.
Here we investigate to see whether smaller-scale neutral vortices can have a higher rotation rate, and with it weaker coupling between the two fluids.

\begin{figure*}
  \centerline{
    \includegraphics[width=15cm]{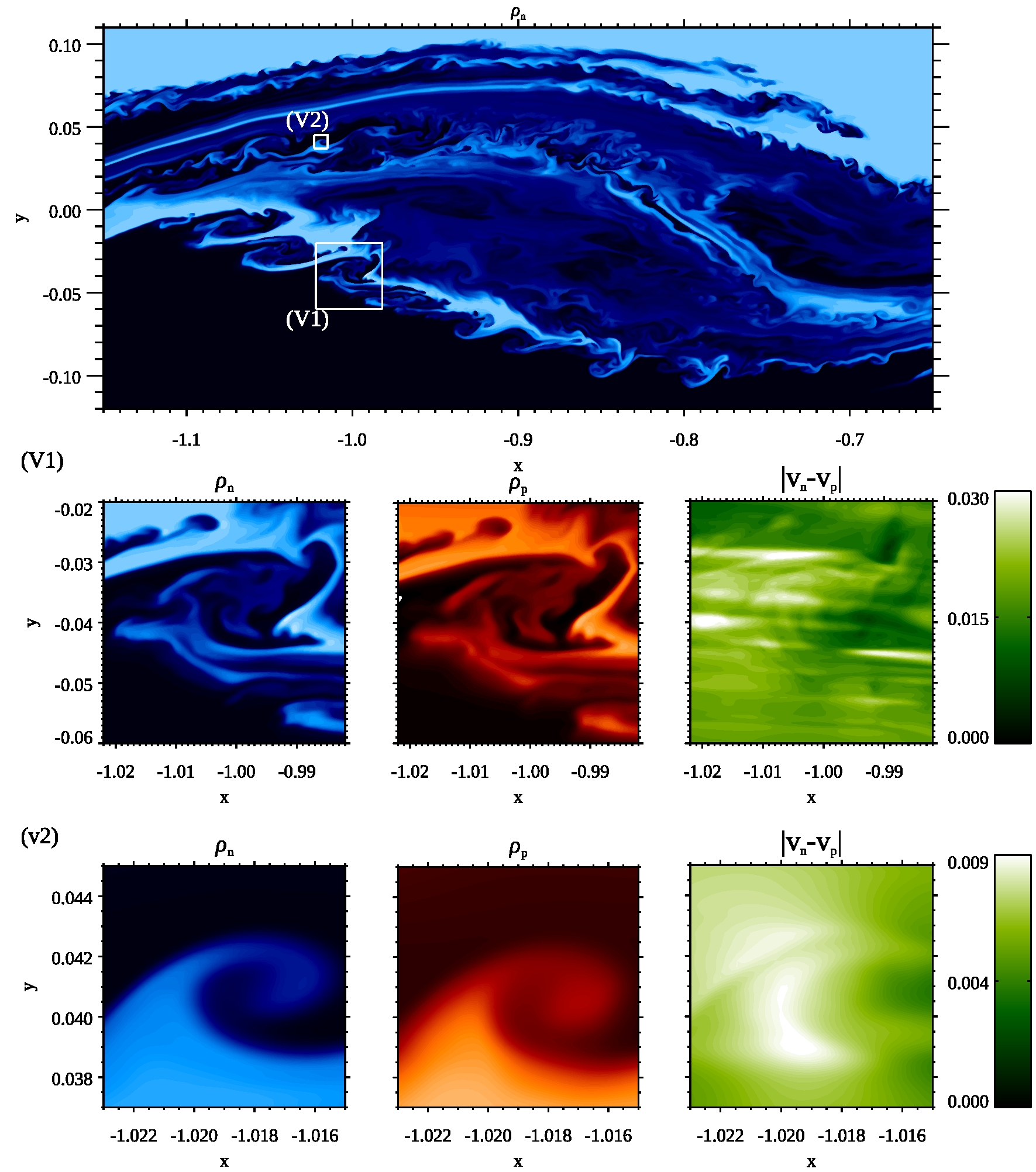}
    }
  \caption{Multiple vortices made through the cascade process from a large vortex {seen at $t=19$}. In the top panel, showing a large scale vortex in the neutral density two vortices (marked V1 and V2 for vortex 1 and vortex 2 respectively) have been isolated. The middle and bottom rows show the neutral and plasma densities, and the drift velocity for these two cases.}
\label{fractal}
\end{figure*}

Figure \ref{fractal} shows a vortex that has formed at $t=19$. 
The top panel of shows the neutral density. In this panel we have isolated an example of a large scale vortex  that has created a number of smaller vortices over a range of scales.
Two particular vortices are highlighted (marked V1 and V2 in the top panel for vortex 1 and vortex 2 respectively) with vortex 2 being a factor of $\sim50$ smaller than the large-scale vortex.

Vortex 1 is less than a tenth of the size of the main vortex. It shows some of the features of the larger vortex, including parasitic Kelvin-Helmholtz and Rayleigh-Taylor instabilities growing on this secondary vortex. The plasma density and neutral density show similar structure. However, the vortex itself does not show up clearly in the drift velocity

Vortex 2 is close to the limit where structures can be resolved (with only $\sim 30$ grid points across it).
The vortex as seen in the plasma density is more diffuse than that of the neutrals. At these scales no substructure is made in the vortex, but we are seeing a clear signal of the vortex rotation in the drift velocity.

\begin{figure*}
  \centerline{
    \includegraphics[width=16cm]{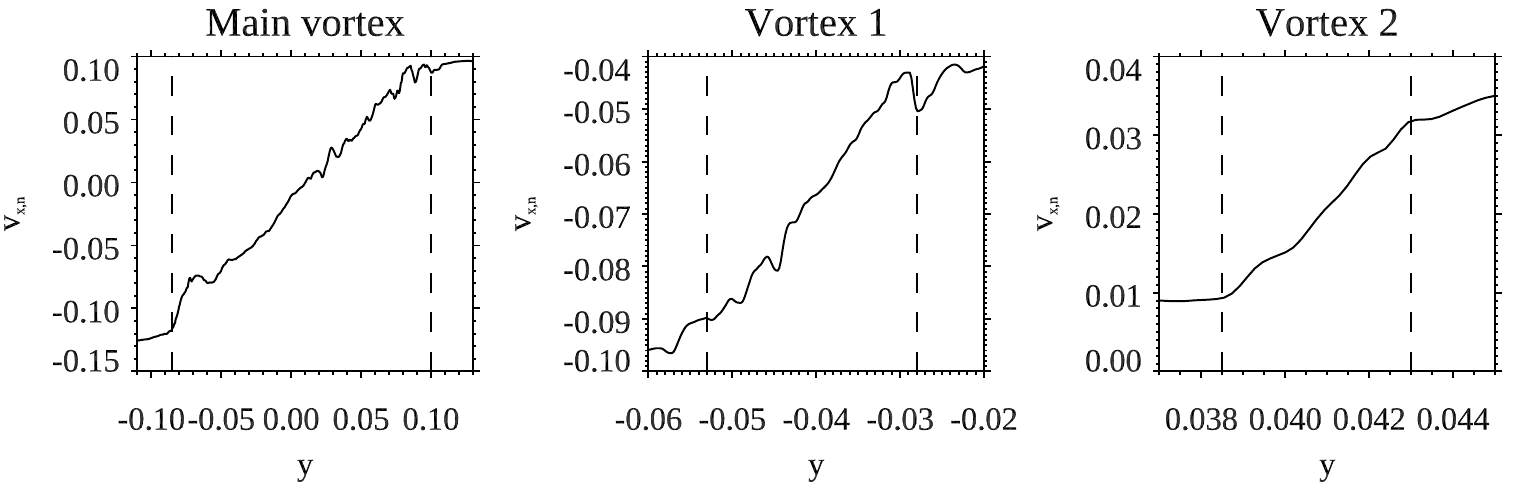}
    }
  \caption{Slice in the $y$-direction of the $v_{\rm x}$ component of the neutral velocity across the different vortices {presented in Figure \ref{fractal}}. The vertical dashed lines show the approximate extent of the vortex as estimated from the density structure. }
\label{shear}
\end{figure*}

Figure \ref{shear} shows the plot of the neutral $v_{\rm x}$ against $y$ across each vortex. These slices are taken over the respective $y$ ranges in the Figure \ref{fractal} plots for $v_{x,n}$ and are are taken at $x=-0.898590$, $x=-0.999847$ and $x=-1.01797$, respectively.
The vertical dashed lines show the boundaries of the vortex. A clear velocity gradient is seen in each case.

The slopes of the velocities shown in these plots are roughly linear. The gradient, a measure of the vorticity, is $\sim1.1$ for the main vortex, $\sim 1.8$ for vortex 1 and $\sim 5.1$ for vortex 2. If the vortices were all rotating at the same rate, then these numbers would be the same, so we are seeing the increase in rotation rate at smaller scales. 

One can roughly estimate how we expect this vorticity to change for Kolmogorov turbulence
The structure function for Kolmogorov turbulence is
\begin{equation}
    \delta v_r\propto r^{1/3}.
\end{equation}
Though this is a statistical property of Kolmogorov turbulence, we use this to approximate the expected velocity difference across the vortex for different scale vortices. This implies that the vorticity measured in each vortex should scale as $r^{-2/3}$. This predicts a vorticity of $12.9$ for the smallest vortex (based on the vorticity of the largest vortex). Therefore we are seeing a system that is somewhere in between these two regimes, but that is exhibiting evidence of an energy cascade through the presence of small scales and greater vorticity at those scales.

As the rotation rate of the vortex is increasing as the scale gets smaller, this implies that at some scale the neutrals will decouple from the plasma. The coupling frequency for the neutrals to the plasma, and with that the frequency to couple to magnetic field, for the parameters of this simulation is $4.5$ (see Section \ref{phys_mean}). Therefore, at the scales of the smallest analysed we are seeing that the neutral fluid motions have become fast enough to decouple.

\subsubsection{Power spectra}

Investigating the power spectra of the velocity and magnetic fields can be a powerful tool for understanding the scales at which energy is held in these fields from a statistical point of view. 
From our reference cases, and the evaluation of the dynamics in our simulation presented previously, we would expect that at scales where the neutral fluid is not coupled to the magnetic field, it will have a spectra similar to that of the HD reference case. However,  if the scales were to get sufficiently large so that coupling of the neutrals to the magnetic field becomes strong, then spectra closer to those of the reference MHD calculation may appear.


\begin{figure*}
  \centerline{
    \includegraphics[width=6.3cm, trim=1cm 0 0.3cm 0, clip]{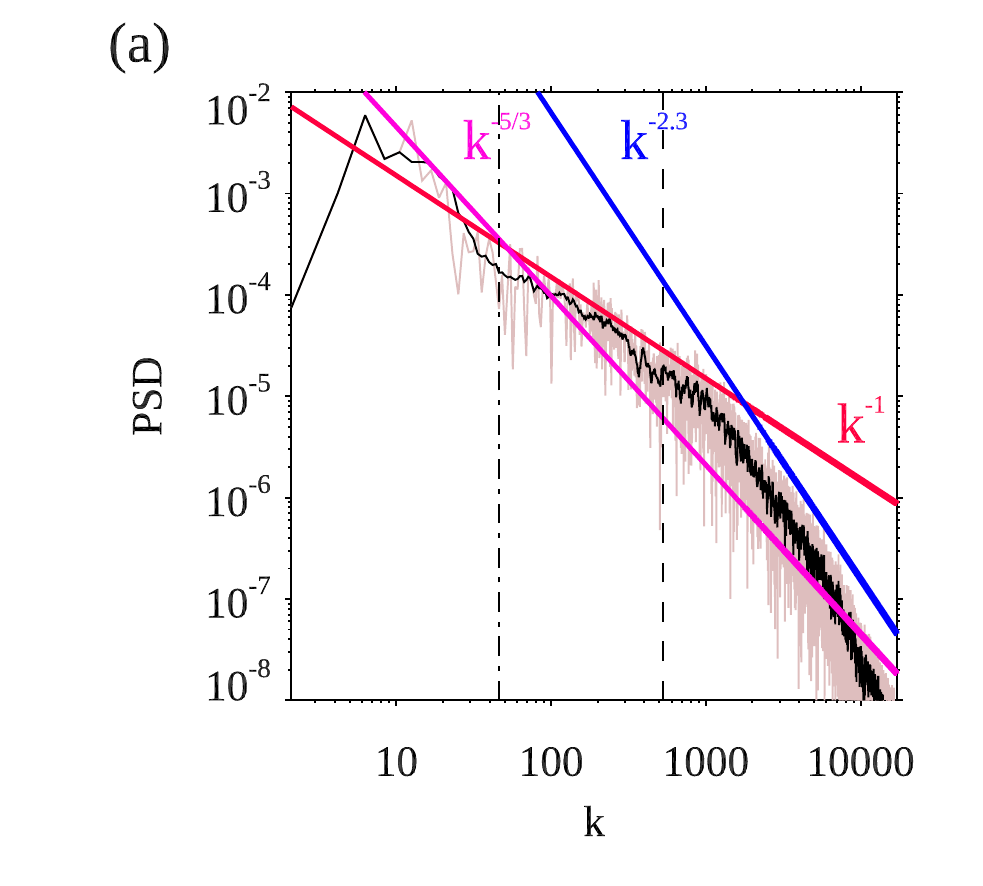}
    \includegraphics[width=6.3cm, trim=1cm 0 0.3cm 0, clip]{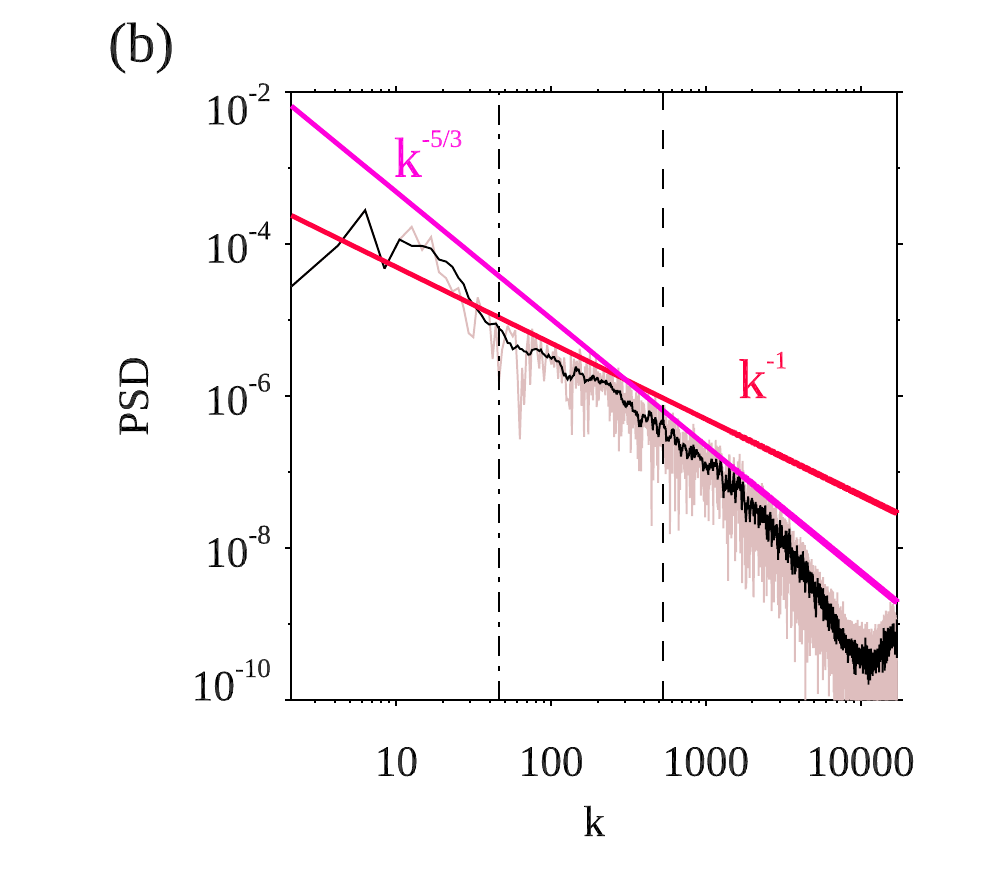}
    \includegraphics[width=6.3cm, trim=1cm 0 0.3cm 0, clip]{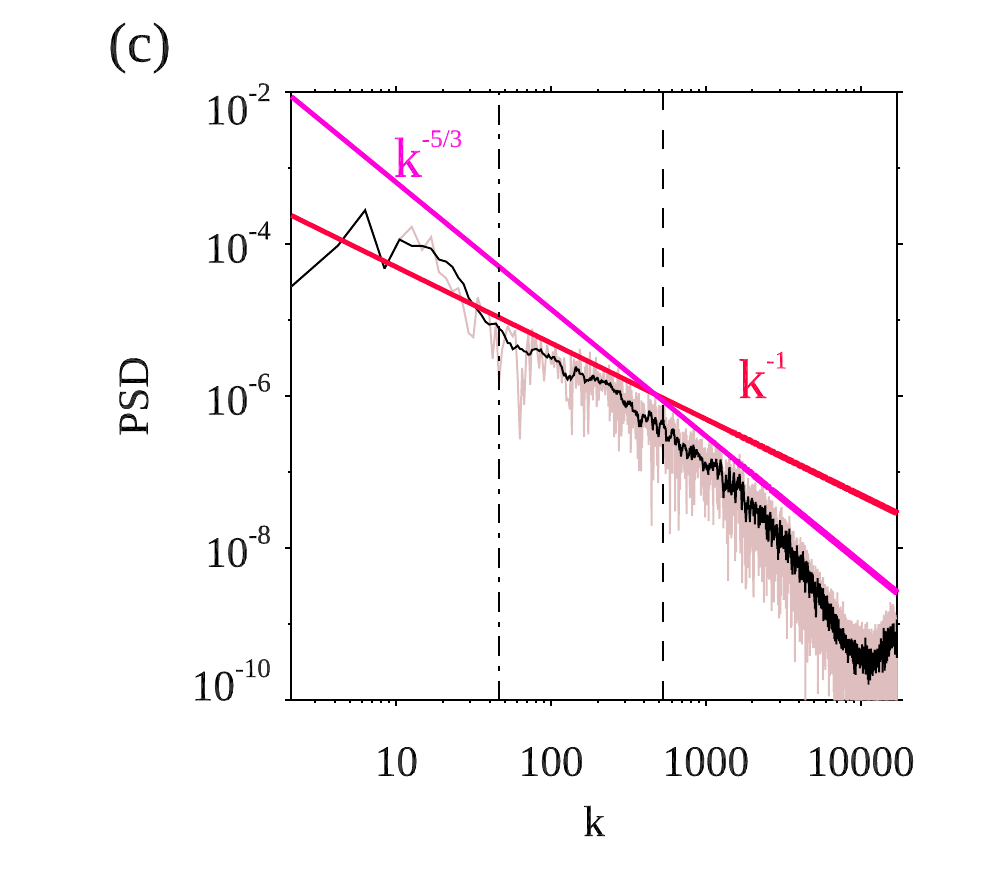}}
  \caption{Power spectra for the density weighted neutral y velocity (panel a), density-weighted plasma y velocity (b) and the y component of the magnetic field at $x=0$ (c) {all taken at $t=19$}.     
  with the black line showing the smoothed spectra. The straight lines show the slopes of $k$ to different exponents used to highlight regions of different behaviour of the different spectra. The vertical dot-dash and dashed line show the approximate scale for neutral and plasma decoupling respectively (see Section \ref{phys_mean}).}
\label{spectra_fig}
\end{figure*}

Figure \ref{spectra_fig} shows a snapshot of the dynamic spectra taken at $t=19$ at $y=0$ for $\sqrt{\rho_n}v_{\rm yn}$ (a), $\sqrt{\rho_p}v_{\rm yp}$ (b), and $B_{\rm y}$ (c).
As these are taken during the dynamic evolution, and not once a statistical steady state has formed, it is not surprising that there is no single slope that dominates the whole of the spectra.
As with the HD reference case, the density weighted neutral velocity (shown in panel a) displays three regions. At the largest scales this appears to be a slope that is greater than $k^{-1}$, with an estimate of $k^{-5/3}$ given {\rm possibly fit these}.
However, the main regions of the spectra consist of an extended region consistent with a power law with exponent of $k^{-1}$ and at higher $k$ approximately following $k^{-2.3}$.
For $k$ larger than $\sim 300$ the plasma velocity spectra and magnetic field spectra diverges from that of the neutral velocity with a spectra of $k^{-5/3}$.

The two vertical lines shown in all three panels give approximate scales at which the neutral decouple from the plasma and with it the magnetic field (dot-dash line) and the scales at which the plasma decouples from the neutral fluid (dashed line), justification of these positions is given in Section \ref{phys_mean}. Above the dashed line, the spectra behave in similar ways, but as this scale is reached the density weighted neutral velocity displays different behaviour, highlighting a dynamic decoupling at small scales.

\subsubsection{frictional heating rate}

As well as momentum transfer, the velocity drift term can provide dissipation in a system.
The frictional heating rate is a second-order term in the velocity difference given by:
\begin{equation}
    H_{\rm FRIC}=\alpha_{\rm c}\rho_n\rho_p\mathbf{v}_{\rm D}^2,
\end{equation}
which determines the dissipation of energy as a result of ion-neutral drift, where $\mathbf{v}_{\rm D}=\mathbf{v}_n-\mathbf{v}_p$.
It is worth noting that these heating terms are different from the standard dissipation terms used in studies of turbulent MHD flows, because unlike Ohmic or viscous heating, they are not dependent on a second derivative, and do not require small scales in either the current or flow to be maximised. This means that to have large heating in a volume it is most effective to have volume filling structures that have a large velocity drift.

\begin{figure}
  \centerline{
    \includegraphics[width=10cm]{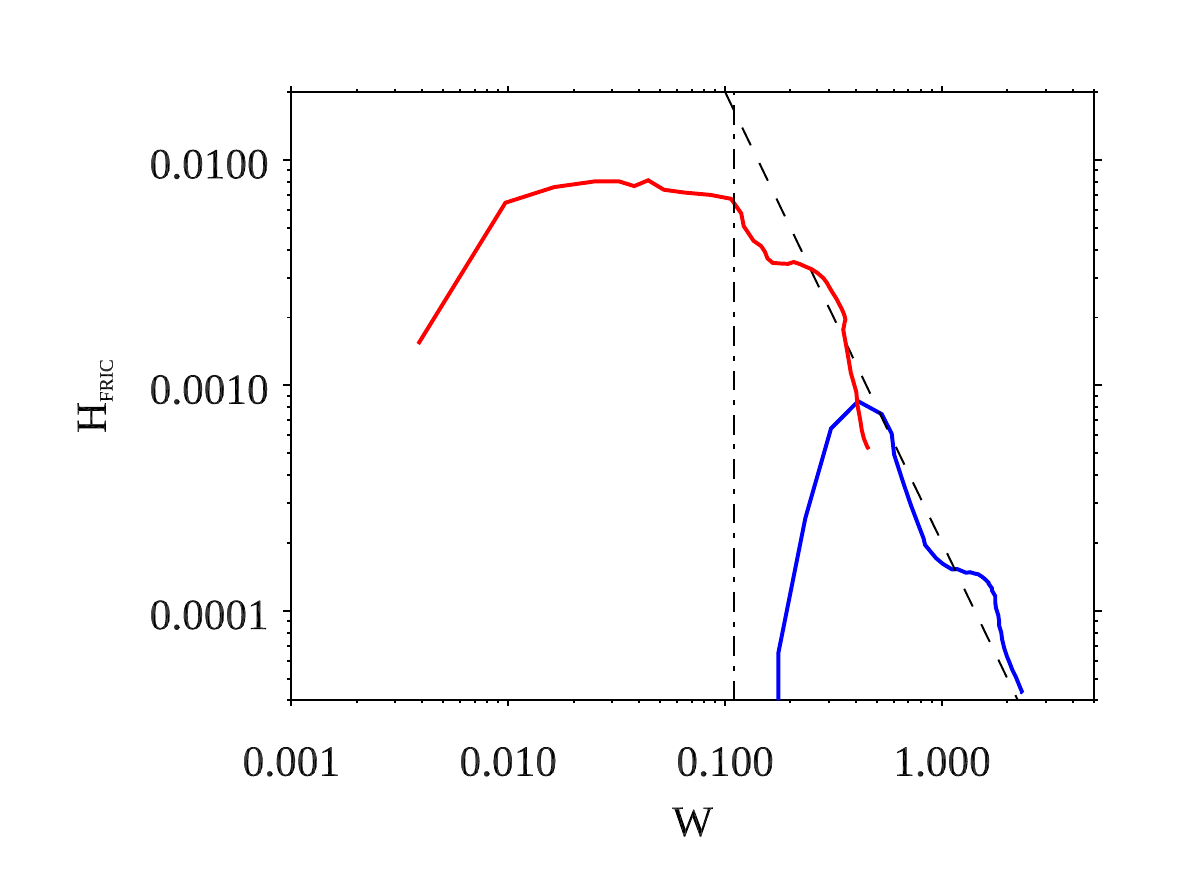}}
  \caption{The frictional heating rate as a function of the width of the KHi region $W$. The red line is calculated for the high resolution simulation, and the blue line is for the simulation performed at larger spatial scales that is explained in Section \ref{large_scale}. The dashed line shows a $W^{-2}$ dependence and the dash-dot line shows the approximate width where the neutral fluid should begin to couple to the magnetic field.}
\label{heat_fig}
\end{figure}

Figure \ref{heat_fig} gives the heating rate for different widths of the KHi layer ($W$) calculated for this simulation, and a simulation that looks at more coupled scales presented in Section \ref{large_scale}.
Here $W$ is calculated by first measuring the minimum and maximum $y$ positions where the $x$ averaged neutral density ($\langle \rho_{\rm n}\rangle$) departs from its initial values either side of $y=0$. Once this initial width has been calculated, it is then widened to capture all the flow dynamics the instability is creating by adding half that width again on both sides of the layer.
The plot shows that at small widths the heating rate has an approximately constant value with width, but above $W\approx0.1$ the heating rate becomes smaller as $W$ increases.

At the smaller values of the width of the mixing layer, the heating rate is saturated at $\sim 0.01$.
An upper bound for the magnitude of the frictional heating can be given by $\alpha\rho_n\rho_p\Delta V^2\approx 0.18$. This means that the actual heating rate is approximately 6 per cent of the upper bound, which is equivalent to the heating region being filled with velocity drifts that average at $\sim \Delta V/5$.

As the width of the mixing layer increases, the system reaches a point where the neutral fluid dynamics begin to couple to the magnetic field (approximately given by the vertical dash-dot line calculated using similar assumptions as those used in the power spectra).
Once coupling begins to take place, this results in a decrease in the drift velocity, and as a result the heating rate starts to decrease as the layer width increases.
Based on arguments of strong coupling of the fluids, then it can be expected that the drift velocity would scale with the Lorentz force, i.e.: 
\begin{equation}
    \alpha_{\rm c}\rho_n\rho_p\mathbf{v}_{\rm D}=(\nabla\times\mathbf{B})\times\mathbf{B}.
\end{equation}
Therefore, we can estimate the heating rate in this regime to be:
\begin{equation}\label{amb_heating}
    H_{FRIC}=\frac{((\nabla\times\mathbf{B})\times\mathbf{B})^2}{\alpha_{\rm c}\rho_n\rho_p}\sim \frac{B^4}{(W/C)^2}\frac{1}{\alpha\rho_n\rho_p}.
\end{equation}
where C is a constant that is used to approximate the magnitude of the current.
This implies that at sufficiently large scales the heating rate can be estimated to scale as $H_{\rm FRIC}\propto W^{-2}$.
The dashed line in Figure \ref{heat_fig} shows the approximate solution in Equation \ref{amb_heating} using the values from the simulation and a constant $C=5$ (note that the choice of $C$ will be heavily dependent on the definition used for $W$).
As can be seen this provides a reasonable representation of the heating rate at those scales. 
This highlights the usefulness of the strong-coupling approximation at large scales, but the danger of applying it to dynamics that do not obey that approximation as heating will be overestimated.

At late times in the high resolution simulation many small scales have been created (e.g. see Fig. \ref{fractal}). Though the rotation rate of the vorticies at the small scales are larger than those of the large-scale vortex driving them, they do not generate large drift velocity. Therefore, their influence on the efficiency of the dissipation is reduced.
Turbulent dissipation through frictional heating can be discussed in a statistical sense using $k$ space, power spectral density ($E(k)$) and the dissipation spectra ($D(k)$) where
\begin{equation}
    \frac{\partial}{\partial t}E(k)=D(k).
\end{equation}
As such we can understand the change in total kinetic energy of the system with time as a result of the frictional heating to be given by:
\begin{align}
    \frac{\partial K}{\partial t}_{\rm FRIC}=\int_0^{\infty}\frac{\partial}{\partial t}E(k)\mathrm{d}k&=\int_0^{\infty}D(k)\mathrm{d}k \nonumber\\
    &\propto \int_0^{\infty}H_{\rm FRIC}(k)\mathrm{d}k.
\end{align}
If we assume for sake of argument that $E(k)\propto k^{-5/3}$ (i.e. is consistent with Kolmogorov turbulence) then $D(k)=\alpha \rho_{\rm n}\rho_{\rm p} v_{\rm D}^2\propto k^{-5/3}$, so even integrating over $k$ space will lead to a dependence with $k$ that has an exponent less than zero.
Therefore we expect that the dissipation from frictional heating will be dominated by the largest scales in the system (unlike viscous or resistive dissipation). 
This explains why even in the presence of an energy cascade, the estimates for the heating in the strong coupling limit given in Equation \ref{amb_heating} prove to be quite accurate even though they only take into account the largest scales of the dynamics.

\subsection{Coupling at even larger scales}\label{large_scale}

To look at larger scales, we have performed a second two-fluid simulation, with resolution of $2048$ by $1024$ over a range of $x=[-15,15]$ and $y=[-7.5,7.5]$.
This means that though we are not able to resolve the decoupled scales, we can look at the way neutral fluids slip across the magnetic field at scales where the neutral fluid can dynamically feel the magnetic forces (i.e. we are in a strong coupling regime).

\begin{figure*}
  \centerline{
    \includegraphics[width=16cm]{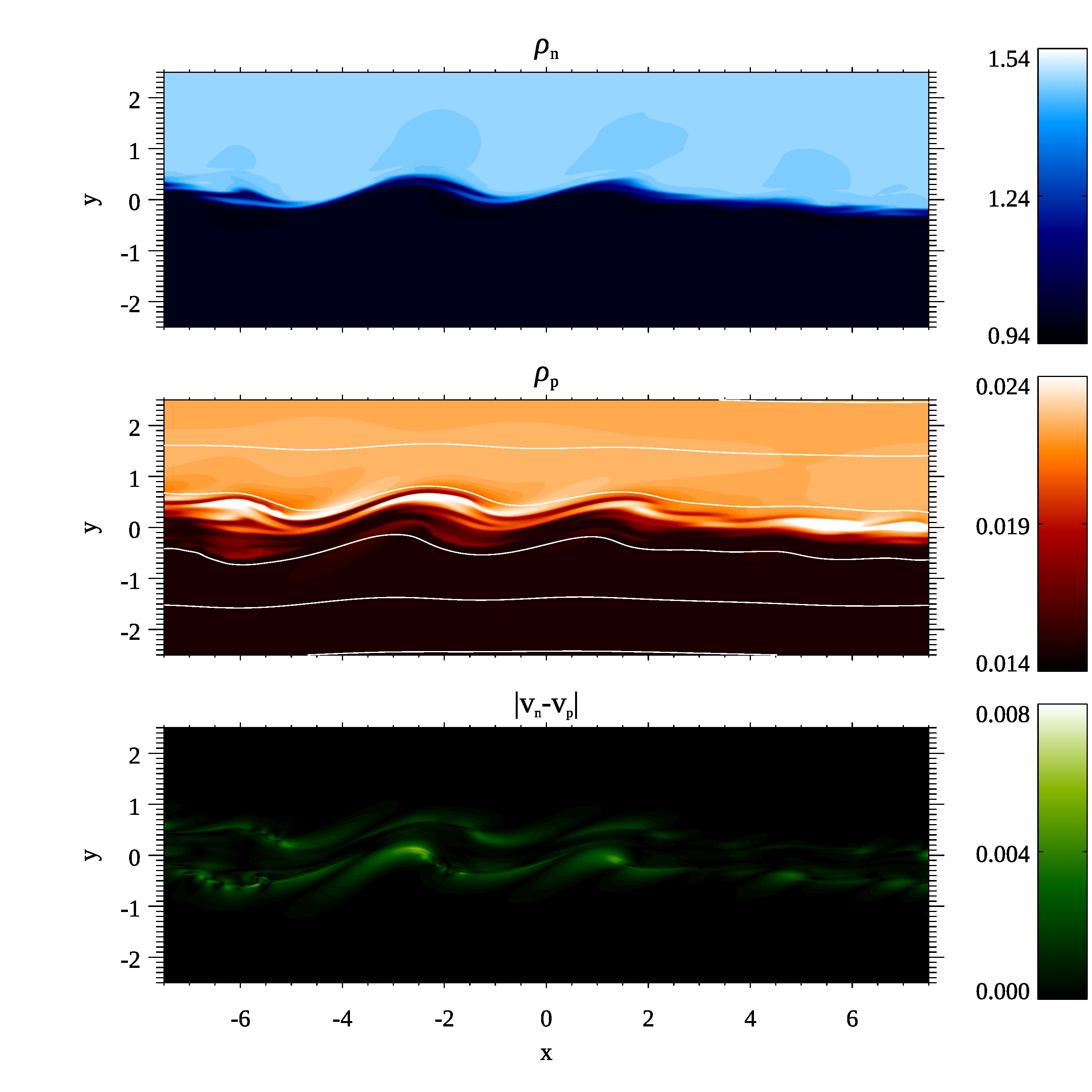}}
  \caption{Snapshot of the simulated densities (top panel for $\rho_{\rm n}$ and middle panel for $\rho_{\rm p}$) and magnitude of the drift velocity (bottom panel) at $t=150$. The white lines in the middle panel show the magnetic field lines. }
\label{t_long_fig}
\end{figure*}

Figure \ref{t_long_fig} shows a snapshot at $t=150$ for the neutral and plasma densities (with fieldlines in white), and the the magnitude of the drift velocity.
One key difference between this and the simulations showing smaller scales is that vortices have not formed in the neutral fluid and overall the simulation result looks closer to that of the reference MHD solution than the reference HD solution.
The neutral density shows striations in the density distribution as a result of drift across the magnetic field.
Though there are similarities between the neutral and plasma density distributions, there is more structure visible in the plasma density distribution.
The drift velocity distribution also shows an interesting feature, the peak velocity drift is not at the interface between the two layers but now there are two regions of high drift just above and below this interface.

The striations of the neutral density emanate from the peaks and troughs of the interface, where the neutral fluid appears to overshoot.
They occur because as the fluid on top is coming out of a furrow or the density on the bottom is coming down from a ridge of the interface, hydrodynamically the fluid wants to roll these structures over to form a vortex, but the magnetic field is resisting. As it is a magnetic force that is suppressing this motion, the neutral fluid (which is not perfectly coupled to the magnetic field), slips across the field lines at these points and intrudes into the the other density layer as a thin strip.

It is ion-neutral drift that creates the neutral density structure, but this is not where the peak drift occurs.
In the reference MHD simulation, the largest deformation of the magnetic field occurred above and below the interface, so in our strongly coupled two-fluid system, this is naturally the place where the largest velocity drift occurs. 
As the neutral density is constant in these regions, and the instability is only weakly compressible, this cannot be seen in the neutral density.
For the component of the flow along the magnetic field the plasma can more follow the neutral fluid, but not across the magnetic field. This creates compressible motions in the plasma creating density changes via compressional field-aligned flows in regions of high neutral drift.

\begin{figure*}
  \centerline{
    \includegraphics[width=6.3cm, trim=1cm 0 0.3cm 0, clip]{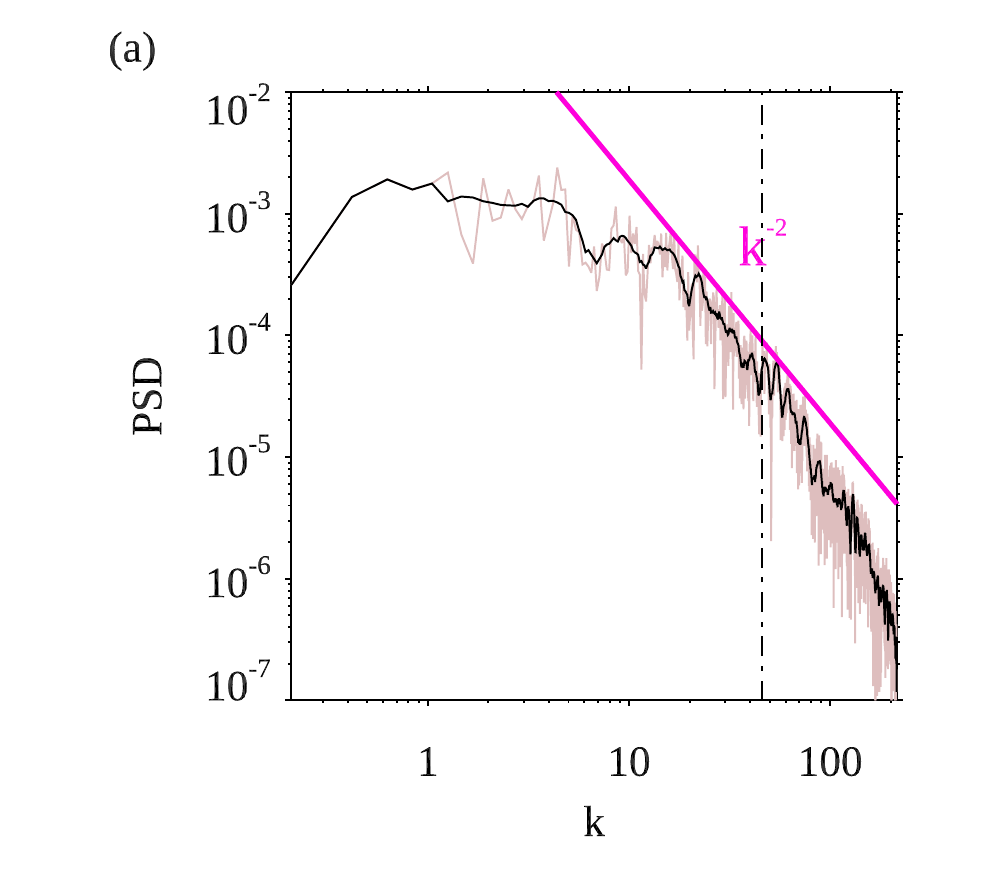}
    \includegraphics[width=6.3cm, trim=1cm 0 0.3cm 0, clip]{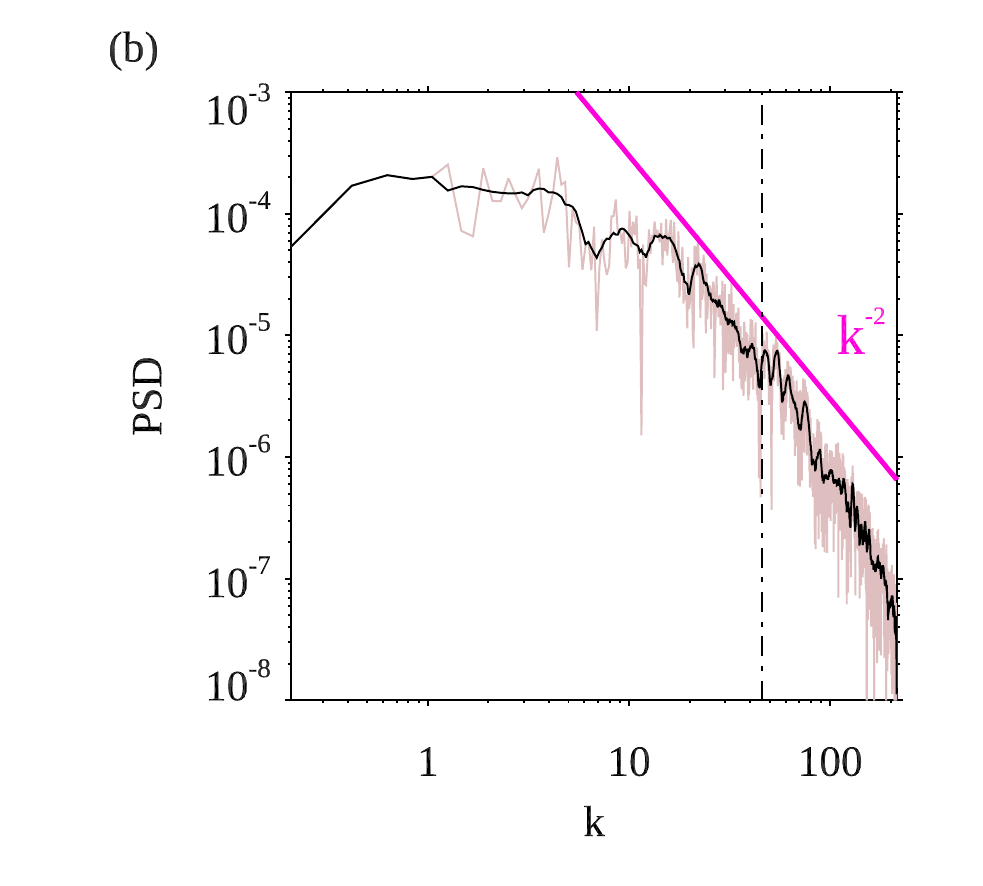}
    \includegraphics[width=6.3cm, trim=1cm 0 0.3cm 0, clip]{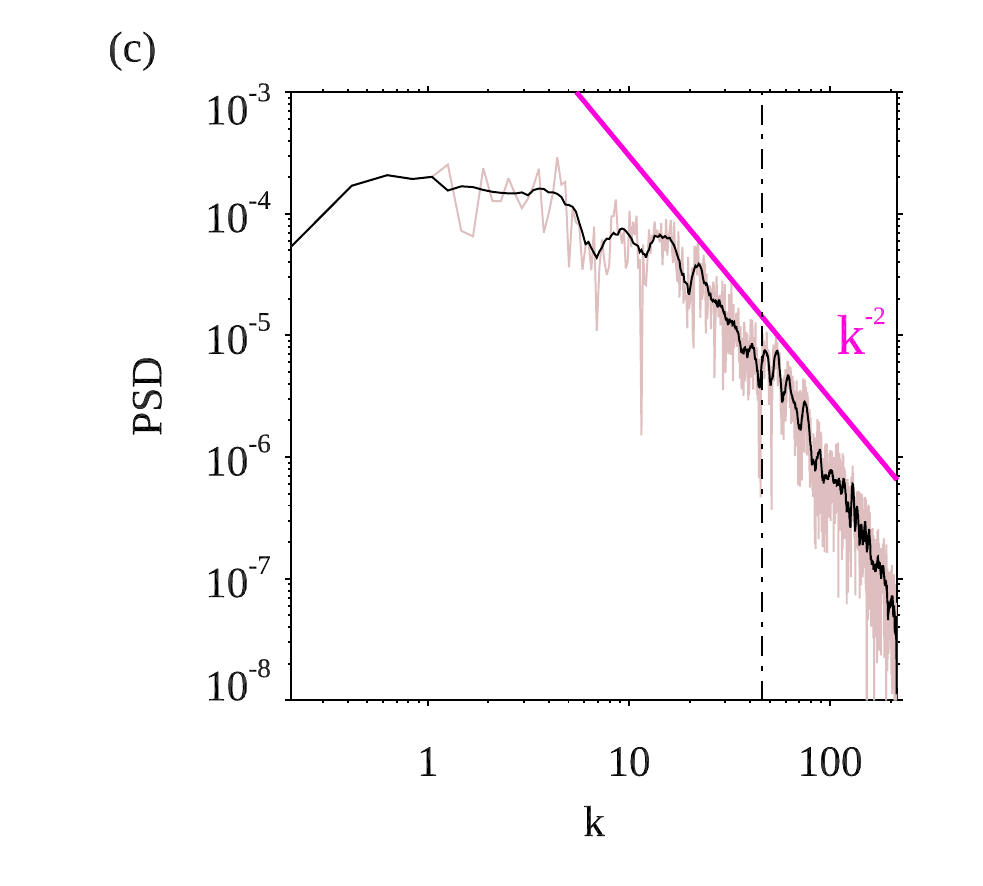}}
  \caption{Power spectra taken at $y=0$ for the density weighted neutral y velocity (panel a), density-weighted plasma y velocity (b) and the y component of the magnetic field (c) {at $t=150$}.     
  with the black line showing the smoothed spectra. The straight lines show the slopes of $k$ to different exponents used to highlight regions of different behaviour of the different spectra.}
\label{long_spectra_fig}
\end{figure*}

Figure \ref{long_spectra_fig} shows the density-weighted $y$ velocity component for the neutrals (panel a) and the plasma (panel b), as well as $B_y$ (panel c).
The slope of the spectra is flat over many wavelengths (as the instability hasn't reached the stage where the largest mode is dominating). At smaller scales a slope approximately consistent with $k^{-2}$ is present. Though this is only clear over a smaller region of the spectra, this slope is consistent with what was seen in the reference MHD solution.
The magnitudes in the power spectral density for the density weighted neutral $y$ velocity component are an order of magnitude larger than those found for the density weighted plasma $y$ velocity and the $B_y$ magnetic field.
The dot-dash line shows an approximation of the scale the neutrals will decouple from the magnetic field, but no change in the spectra is obvious around this region.

\section{Conclusions}

In this paper we have investigated the interaction between the neutral and
plasma fluids in the nonlinear stage of the Kelvin--Helmholtz instability. 
What we have found is that: if dynamics are sufficiently high frequency then all the complex motions are occurring in the neutral fluid. 
However, as the instability dynamics occurs at progressively larger scales, the fluids became progressively more coupled together. 
{This increase in coupling results in the magnetic field being able to indirectly exert a force on the neutral fluid, resulting in a flattening of the vortices.
Once the large-scale vortices form, as part of the energy cascade that develops, motions at decoupled scales are also driven in the system.}

At scales where the neutral dynamics are decoupled from the magnetic field, fully-formed neutral vortices are present. The plasma velocity field does not have this response.
Though the velocity structure decouples, the density structures of the two fluids is coupled for dynamic frequencies below the ion-neutral collision frequency as a result of the heating and cooling of the plasma by the thermal terms in the energy equations. For higher frequency dynamics not only the flow but also the temperature become decoupled removing this effect.
This density coupling effect can be shown to occur even for very low plasma $\beta$ as it partly involves compression and expansion along the magnetic field.
At the scales where the neutral flow is decoupled from the magnetic field, the frictional heating is independent of the mixing layer thickness. Above that scale the heating scales as $W^{-2}$. This highlights how ambipolar diffusion would overestimate heating rates for dynamics below the strong coupling scale.

Even at the point where large scale vortices have formed and the neutral fluid is beginning to couple with the magnetic field, the smaller vortices they produce can become decoupled. 
This means that at as part of a KHi turbulence energy cascade, when scales where the neutrals have decoupled from the magnetic field are reached, fully-developed, purely-hydrodynamic turbulence can exist and contain the majority of the energy of the cascade.
One important consequence of this work is that in an energy cascade in MHD, when motions reach these high frequencies it is likely that a transition from MHD turbulence in the coupled fluids to the neutral fluid possessing its own turbulent cascade will develop.
In terms of the dissipation effects in two-fluids, however, it will always be the larger scales that contribute more, meaning that the development of a purely neutral fluid energy cascade will require viscosity to dissipate the energy in the neutral cascade.

An important corollary that comes from this work, applicable equally to purely MHD systems and PIP systems, is to do with the role of magnetic fields in the nonlinear saturation of the instability.
In terms of having this instability be dynamically important, it is necessary to go beyond whether the system is linearly unstable, and look at how the nonlinear saturation of the instability will behave at the scales of interest.
This has been seen in previous studies\citep{RYU2000}, and here we have extended this idea by showing that the linear stability problem can be used as a good tool to estimate the nonlinear regime the simulation enters.

\vskip 0.5cm

Andrew Hillier is supported by his STFC Ernest Rutherford Fellowship grant number ST/L00397X/2 and STFC research grant ST/R000891/1. This work used the COSMA Data Centric system at Durham University, operated by the Institute for Computational Cosmology on behalf of the STFC DiRAC HPC Facility (www.dirac.ac.uk). This equipment was funded by a BIS National E-infrastructure capital grant ST/K00042X/1, DiRAC Operations grant ST/K003267/1 and Durham University. DiRAC is part of the UK National E-Infrastructure. AH would like to thank Drs. Shinsuke Takasao and Naoki Nakamura for their work on developing the (P\underbar{I}P) code, and their support and advice in this study. AH would also like to thank Drs. Roberto Soler and Benjamin Snow for their comments on the manuscript.
\nocite{*}
\bibliography{PIP_ref}

\end{document}